# A *d*-Electron Heavy-Fermion-Like Superconductor with Frustration-Induced Flat Bands

Chaoguo Wang,[1†] Hyeonhu Bae,[2†] Jiaqi Tian,[1†] Qing-Ping Ding,[3] Yongbin Lee,[3] Yuji Furukawa,[3,4] Shannon Gould,[5] Brianna R. Billingsley,[6] Si Athena Chen,[7] Matthias Frontzek,[7] Tai Kong,[6,8] Binghai Yan,[2,9] Sheng Ran[5] and Xin Gui[1]*

[1] Department of Chemistry, University of Pittsburgh, Pittsburgh, PA, 15260, USA
[2] Department of Condensed Matter Physics, Weizmann Institute of Science, Rehovot, 7610001, Israel
[3] Ames National Laboratory, U.S. Department of Energy, Ames, IA, 50011, USA
[4] Department of Physics and Astronomy, Iowa State University, Ames, IA, 50011, USA
[5] Department of Physics, Washington University at St. Louis, St. Louis, MO, 63130, USA
[6] Department of Physics, University of Arizona, Tucson, AZ, 85721, USA
[7] Neutron Scattering Division, Oak Ridge National Laboratory, Oak Ridge, TN, 37831, USA
[8] Department of Chemistry and Biochemistry, University of Arizona, Tucson, AZ, 85721, USA
[9] Department of Physics, The Pennsylvania State University, University Park, PA, 16802, USA

[†]These authors contributed equally to this work.

*Correspondence to: xig75@pitt.edu

## Abstract

**Heavy-fermion superconductors are mostly associated with *f*-electron materials with Kondo lattices, while known *d*-electron heavy-fermion-like systems are often linked to orbital-selective local moments, Hund-metal physics, or a charge-density-wave mechanism. Here we report $Mo_4PtGa_{17}$, a noncentrosymmetric itinerant *d*-electron superconductor with a geometrically frustrated breathing-pyrochlore Mo lattice. Thermodynamic, transport and NMR measurements reveal heavy-fermion-like behavior superconductivity and dominant ferromagnetic spin fluctuations near a ferromagnetic instability. Theoretical calculations identify nearly flat bands, van Hove singularities and Kramers nodal lines near the Fermi energy, derived intrinsically from Mo-4*d* states and are robust against on-site electronic correlations. These results suggest that the geometrically frustrated lattice in $Mo_4PtGa_{17}$ generates an intriguing electronic structure that enhances the density of states, spin susceptibility and quasiparticle mass. $Mo_4PtGa_{17}$ therefore** **identifies a unique route to heavy-fermion-like superconductivity in *d*-electron materials through geometrical frustration, different from the previously reported systems.**

Strong electronic correlations can suppress conventional phonon-mediated pairing while enabling unconventional superconductivity, including cuprates[1], iron-based superconductors,[2] moiré materials[3,4] and heavy-fermion superconductors (HFSCs)[5–8]. Among these, HFSCs occupy a distinct regime where superconductivity emerges from quasiparticles whose masses are strongly enhanced by many-body correlations, most realized in *f*-electron Kondo lattices.[5] Extending this physics beyond *f*-electron materials is an important challenge because it could reveal new mechanisms for generating heavy-quasiparticle superconductivity from more itinerant electronic states. A few *d*-electron materials demonstrated that heavy-quasiparticle behavior can arise without *f* orbitals.[9–11] However, these systems are generally associated with microscopic routes specific to each material system: orbital-selective local moments in $LiV_2O_4$, which is non-superconducting,[12] Hund-metal/orbital-selective physics in iron-based superconductors,[13,14] and charge-density-wave-related mechanism in nickel chalcogenides superconductors.[15] Identifying distinct routes to *d*-electron heavy-fermion-like superconductivity will offer novel material design principles and it remains a central materials problem.

Geometrical frustration offers a unique possibility, where destructive interference between hopping pathways can generate weakly dispersive bands and van-Hove singularities (vHS) near the Fermi level.[16,17] These features can enhance the electronic density of states, spin susceptibility and quasiparticle mass while preserving itinerant nature. Kagome superconductors, such as $AV_3Sb_5$ (A = K, Rb, Cs), showcase how frustrated *d*-electron lattices can host intertwined electronic orders and superconductivity, but their mass enhancement remains moderate for heavy-fermion-like regime.[18–20] This leaves open the possibility of utilizing geometrical frustration to drive heavy-fermion-like superconductivity in itinerant *d*-electron materials.

Here, we report the discovery and characterizations of $Mo_4PtGa_{17}$, a new itinerant *d*-electron superconductor hosting heavy-fermion-like behavior, enhanced ferromagnetic spin fluctuations, nearly flat bands, vHS and nodal lines, while resembling a Kramers nodal line metal, making it a topological nodal line metal candidate. Geometrically frustrated Mo breathing pyrochlore lattice is found to play a crucial role in hosting the intriguing electronic structure and inducing the exotic quantum states. Our discovery opens a new route to realizing heavy-fermion-like superconductivity in *d*-electron systems, distinct from *f*-electron Kondo lattices, iron-based Hund metals and charge-density-wave *d*-electron systems. It also provides a novel avenue for designing correlated superconductors from geometrically frustrated itinerant ferromagnets, given its ferromagnetic isostructural analogue, $Cr_4PtGa_{17}$.[21]

**Geometrical Frustration and Phase Purity:** $Mo_4PtGa_{17}$ crystallizes in the noncentrosymmetric trigonal space group, *R3m* (No. 160), see Fig. 1a, isostructural to $Cr_4PtGa_{17}$.[21] Crystallographic data are listed in Extended Data Tables S1-S3. The Flack parameter (0.019 (14)) suggests noncentrosymmetric symmetry. $Mo_4PtGa_{17}$ consists of Mo@$Ga_9$ triply capped trigonal prism (Fig. 1b) embedded within Pt@$Ga_6$ octahedra. Seven Ga sites, one Pt site and two Mo sites, i.e., Mo1 (Wyckoff 9*b*) and Mo2 (Wyckoff 3*a*), are seen. A breathing pyrochlore lattice of Mo is shown in Fig. 1c, with a breathing Kagome sublattice (Mo1) and a triangular sublattice (Mo2) alternately stacking along [001]. The Mo1-Mo2 bond length is ~3.1 Å in the smaller Mo2@$(Mo1)_3$ tetrahedra and ~5.1 Å in the bigger ones, forming $Mo_4$ tetrahedral cluster framework (Fig. 1d). Powder X-ray and neutron diffraction, using HB-2C $WAND^2$ at Oak Ridge National Laboratory, confirm the high purity of the sample, as discussed in Extended Data Fig. S1 and Supplementary Information.

**Magnetic Behavior:** Temperature-dependent magnetic susceptibility, $\chi(T)$, is measured under zero-field cooling (ZFC) and field-cooling (FC) protocols, as seen in Fig. 2a. The high-temperature paramagnetic region is fitted using the Curie-Weiss law, as shown in Supplementary Information. For both ZFC and FC protocols, the fitted $\mu_{eff}$ is ~2.35 $\mu_B$/f.u., i.e., ~0.59 $\mu_B$/Mo, smaller than the lowest expected localized moment, i.e., S = ½ configuration, which is a typical behavior of itinerant materials. The small $\mu_{eff}$ is close to the isostructural ferromagnetic $Cr_4PtGa_{17}$ (~2.0 $\mu_B$/f.u.).[21] Given the itinerant ferromagnetism of $Cr_4PtGa_{17}$ suggested by Rhodes-Wohlfarth ratio (RWR) of ~1.58,[22–24] it is plausible that $Mo_4PtGa_{17}$ also possesses itinerant behavior.

A broad maximum ($\chi_{max}$) is observed for both ZFC and FC curves at ~30 K, while the fitted $\theta_{CW}$ are -47.2 (3) K and -51.9 (3) K for ZFC and FC, respectively. Isothermal magnetization loops were collected and shown in Fig. 2b. Nearly linear loops under all temperatures are observed, indicating the absence of long-range ferromagnetic order and saturation below 9 T. The possibility of spin-glass state is excluded while trace amount of superconducting impurities are seen, which are determined to have negligible impact on the bulk magnetic properties, as seen in Extended Data Figs. S2 and S3 and Supplementary Information.

**Electronic Correlations:** Heat capacity ($C_p$) measurement in Fig. 2c exhibits the absence of long-range magnetic ordering, see Supplementary Information. To better understand the measured $\chi(T)$ and $C_p(T)$, the subtraction of phononic contributions is attempted in the absence of non-magnetic analogues. However, no reliable results are obtained, as seen in Extended Data Fig. S4 and Supplementary Information. To estimate electronic contributions to $C_p$, the low-temperature $C_p(T)$ is fitted and shown in the right inset of Fig. 2c utilizing

$$C_p = \gamma T + \beta T^3 + \alpha T^5$$

where γ, β and α represent the electronic, phononic and anharmonic phononic contributions, respectively. The fitted γ, β and α are 121 (2) mJ/mol/K$^2$, 0.41 (8) mJ/mol/K$^4$ and 0.0087 (8) mJ/mol/K$^6$, respectively. The obtained γ is large compared to normal metals (<10 mJ/mol/K$^2$), e.g., Al,[25] or some transition-metal-based quantum-critical systems including $BaFe_2As_2$ (5 mJ/mol/K$^2$) and derivatives (~25 mJ/mol/K$^2$).[26] In fact, $\gamma(Mo_4PtGa_{17})$ is comparable to many heavy-fermion systems, e.g., $UTe_2$ (110 mJ/mol/K$^2$),[8] $YMn_2$ (180 mJ/mol/K$^2$),[27] and $UAl_2$ (145 mJ/mol/K$^2$),[28] suggesting strong electronic correlations. Moreover, as illustrated in Extended Data Fig. S5, the linear relation of $C_p/T(T^2)$ below ~7 K, i.e., $T^2 \lesssim 49$ K$^2$, reveals insignificant contribution from $C_p/T \propto \alpha T^4$ term under low temperatures. Thus, γ can be visualized by extrapolating the low-temperature linear portion to the y axis. Clearly, magnetic fields up to 9 T do not significantly vary γ above 2 K.

**Ferromagnetic Spin Fluctuations:** Although $C_p(T)$ is rather featureless, it should be noted that there are spin-fluctuating systems adopting similar behavior, e.g., Pd and $Mo_3Sb_7$.[29,30] In fact, by having γ and the magnetic susceptibility for T→0 ($\chi_{T\to 0}$), one can obtain the Wilson ratio, $R_W = \frac{\pi^2 {k_B}^2}{3{\mu_B}^2}\frac{\chi_{T\to 0}}{\gamma}$,[31] revealing the nature of spin correlations. As shown in Fig. 2a, $\chi_{T\to 0}$ can be taken as 0.0062 emu/Oe/mol, resulting in $R_W \sim 3.7$. Given $R_W$(free electron) ~ 1, it suggests that $Mo_4PtGa_{17}$ possesses strong spin correlations and significant spin fluctuations. Moreover, one can calculate the Stoner enhancement parameter using $Z = 1 - \frac{3{\mu_B}^2}{\pi^2 {k_B}^2}\frac{\gamma}{\chi_{T\to 0}}$,[31] which reflects the strength of spin correlations and how close the system is to a ferromagnetic instability. Here, Z($Mo_4PtGa_{17}$) ~ 0.73, comparable to other nearly ferromagnetic Fermi liquids, e.g., $YFe_2Zn_{20}$ (0.88),[32] $LuFe_2Zn_{20}$ (0.89),[32] Pd (0.83),[33] and $YCo_2$ (0.75).[34] This further indicates that $Mo_4PtGa_{17}$ is among the Stoner-enhanced materials with strong spin correlations and fluctuations in the vicinity of ferromagnetism, consistent with its ferromagnetic derivative, $Cr_4PtGa_{17}$.[21]

To confirm the type of spin fluctuations, $^{69,71}$Ga NMR is conducted on polycrystalline samples from 1.6 K to 300 K (see Extended Data Figs. S6 and Supplementary Information for details). As shown in Fig. 2d, temperature dependence of the spin part of Knight shift ($K_{spin}$) of Ga1 site is consistent with that of χ(T) in Fig. 2a where a maximum is observed at ~30 K in χ(T) and a minimum is seen at the same temperature in $K_{spin}$(T) due to the negative hyperfine coupling constant. Additionally, a broad maximum also emerges in $1/T_1T$(T) shown in Fig. 2e. The similar temperature dependence of $K_{spin}$ and $1/T_1T$ suggests predominant ferromagnetic spin fluctuations in nature. In fact, the FM fluctuations can be shown by modified Korringa analysis, $T_1TK_{spin}^2 = \frac{\hbar}{4\pi k_B}\left(\frac{\gamma_e}{\gamma_N}\right)^2 = S$, where $\gamma_e$ and $\gamma_N$ are electron and nuclear gyromagnetic ratios, respectively. Deviations from $S$ can reveal information about magnetic fluctuations in materials,[35,36] and are expressed *via* the parameter $\alpha = S/T_1TK_{spin}^2$, where $\alpha < 1 (> 1)$ corresponds to ferromagnetic (antiferromagnetic) fluctuations.[37,38] As shown in the inset of Fig. 2e, although $\alpha$ is close to unity at high temperatures, it decreases when cooling and levels off below 100 K till 1.6 K at $\alpha \sim 0.15$, evidencing enhanced ferromagnetic fluctuations below ~100 K. It is noteworthy that net antiferromagnetic correlations are observed above 100 K, as suggested by Fig. 2a where χ(T) deviates from Curie-Weiss behavior. Therefore, it is plausible that competing spin correlations exist while high-temperature region is dominated by antiferromagnetic correlations and low-temperature region is dominated by ferromagnetic fluctuations. However, to determine the existence of finite-$q$ fluctuations at low temperatures, a full $q$-resolved map of $\chi(q,\omega)$, i.e., inelastic neutron scattering, is required.

**Heavy-Fermion-like Behavior:** To interpret the electronic properties of $Mo_4PtGa_{17}$, temperature-dependent electrical resistivity (ρ) is measured from 2 K to 300 K under zero magnetic field (Fig. 2f). Due to the synthetic method, multi-domain crystals that naturally grow together are used for measurements. A clear metallic behavior can be seen with a residual-resistance ratio (RRR) of ~ 27, suggesting good crystal quality despite the multi-domain nature. The low-temperature data (< 10 K) is fitted with

$$\rho = \rho_0 + AT^2$$

where $\rho_0$ is residual resistivity and A is a constant reflecting the electron-electron scattering strength. Here, A($Mo_4PtGa_{17}$) is 0.154 (9) μΩ cm/$K^2$, suggesting Fermi-liquid behavior under low temperatures and strongly correlated electrons, consistent with the large γ observed in $C_p(T)$. It is also comparable to other strongly correlated systems[39] including FeSe (0.260 μΩ cm/$K^2$)[40] and several heavy-fermion systems, e.g., $Y_{1-x}Sc_xMn_2$ (0.25 μΩ cm/$K^2$)[41] and $UAl_2$ (0.19 μΩ cm/$K^2$)[42]. Additionally, the Kadowaki-Woods ratio, $R_{KW} = A/\gamma^2$, can be used to describe the relationship between the electron-electron scattering rate and the renormalization of the electron mass. For example, $R_{KW}$ is typically $10^{-5}$ μΩ cm mol$^2$ K$^2$/mJ$^2$ or above in heavy-fermion systems.[39,42] It is found that $R_{KW}(Mo_4PtGa_{17}) \sim 1.05 \times 10^{-5}$ μΩ cm mol$^2$ K$^2$/mJ$^2$, suggesting heavy-fermion-like behavior.

**Superconducting Properties:** Low-temperature ρ(T) is measured from 0.1 K to 1.0 K under various magnetic fields and a current of 100 μA. As shown in the main panel of Fig. 3a, a clear resistance drop is seen under zero field starting ~0.61 K ($T_{c\text{-}onset}$) to zero resistance at ~0.5 K ($T_{c\text{-}zero}$), indicating superconductivity. With higher magnetic field up to 0.5 T, the superconducting transition is gradually suppressed. By taking the corresponding temperature of the transition midpoint under each field, superconducting upper critical field ($\mu_0H_{c2}(0)$) is fitted to be 0.307 (3) T while $T_c$ is 0.566 (4) K using the Ginzburg-Landau relation

$$\mu_0 H_{c2}(T) = \mu_0 H_{c2}(0)\frac{(1-t^2)}{(1+t^2)}$$

where t = T/$T_c$ and $T_c$ is the transition temperature at zero magnetic field, as shown in the inset of Fig. 3a. The superconducting coherence length is estimated to be ~35 nm, as seen in Extended Data Fig. S7 and Supplementary Information. The critical current ($I_c$) is determined by sweeping the applied current under various temperatures. As illustrated in Fig. 3b, clear zero resistance is present at 0.05 K within ~ ±0.85 mA and is suppressed to lower current with increasing temperature regardless of the sweeping direction. Extended Data Fig. 7b demonstrates the temperature-dependence of $I_c$ extracted from Fig. 3b and it gives $I_c$ ~ 0.85 mA at 0.05 K. In addition, AC magnetic measurement is performed below 1 K and is plotted in Fig. 3c. A clear diamagnetic signal suggesting Meissner effect is observed below ~ 0.6 K and saturates at ~ 0.5 K, which confirms the existence of superconductivity.

Heat capacity measurement is conducted between 0.1 K and 2.0 K to confirm bulk superconductivity. As seen in the main panel of Fig. 3d, a λ-anomaly can be observed at ~0.62 K under zero magnetic field, consistent with above. With $\mu_0H$ = 0.4 T, the onset of superconducting transition is suppressed and absent above 0.15 K, consistent with ρ(T). The normalized superconducting jump ($\frac{\Delta C}{\gamma T_c}$) at zero field is ~ 1.09, as shown in the inset of Fig. 3d where the entropy conservation is found, smaller than the value expected for weakly-coupled BCS superconductors, 1.43. By fitting $C_p$(T) below $T_c$ (0.1 K – 0.48 K) using the formula for BCS superconducting gap, $C_p = Ae^{-\Delta/T}$, where A is a constant related to density of states at $E_F$ and Δ is the superconducting gap, the model does not resemble the low-temperature zero-field $C_p$ (upper panel in Fig. 3e). However, a better fitting is obtained with the two-gap BCS relation, $C_p = BA_1e^{-\Delta_1/T} + (1 - B)A_2e^{-\Delta_2/T}$, where $A_1$ and $A_2$ are constants, B is the weighting factor and $\Delta_1$, $\Delta_2$ are the superconducting gap. Here, B is fitted to be ~0.15 while $\Delta_1$ and $\Delta_2$ are ~0.15 (4) K and ~0.94 (3) K, indicating fully gapped superconductivity.

To obtain more information on the superconducting gap symmetry, $^{69}$Ga NQR is conducted on polycrystalline samples across $T_c$. As shown in the main panel of Fig. 3f, Hebel-Slichter (HS) peak can be observed just below $T_c$ in the $1/T_1$ (T) curve, followed by an exponential decrease at lower temperatures fitted by $1/T_1 = Ae^{-\Delta/T}$ where A ~ 9.2 and Δ is ~0.60 (1) K, consistent with a fully gapped state.[43] The gap magnitude of $\Delta/k_BT_c$ ~ 1 is smaller than 1.76, expected from BCS theory, which can be explained by the potential chemical disorders in the powder samples used for NMR measurements that lead to a distribution of $T_c$. The Hebel-Slichter peak can be seen more clearly from the $1/T_1T$ plot shown as the inset of Fig. 3f. Under small magnetic fields of 0.02 T and 0.04 T, the HS peak is suppressed, which is consistent with the behavior of other conventional superconductors. Given the consistent results between heat capacity and NMR measurements, $Mo_4PtGa_{17}$ is likely to be a fully gapped *s*-wave superconductor surviving under a heavy-fermion-like correlation and spin-fluctuated background.

**Electronic Structure:** To understand the electronic properties of $Mo_4PtGa_{17}$, density functional theory (DFT) calculations are performed and focused on the low-energy electronic structure near $E_F$. We observed van Hove singularities (vHS), nodal lines and nearly flat bands near $E_F$, predominantly from Mo-4*d* states. Fig. 4a showed the nonmagnetic metallic band structure with and without spin-orbit coupling (SOC) based on the high-symmetry points shown in Extended Data Fig. S8a. The band structure of $Mo_4PtGa_{17}$ mimics a heavily p-doped semiconductor. Without SOC, several bands near $E_F$ show saddle-point-like behavior, e.g., near L and W points, and lead to a peak of DOS, which is split after including SOC with ΔSOC at Γ ~116 meV. Fig. 4b shows orbital-projected bands for Mo-4*d* and Ga-4*p*. In both non-SOC and SOC calculations, the bands controlling the DOS near $E_F$ are predominantly Mo-4*d* derived, with momentum-dependent Ga-4*p* that becomes enhanced in selected momentum regions with stronger hybridization like Γ. Meanwhile, Pt contributes mainly through deeper-lying 5*d* states around -3.5 eV, and only a negligible weight near $E_F$ from mainly *p*-orbitals. Extended Data Fig. S8b compares SOC band structures for the nonmagnetic and a weakly ferromagnetic phase. The negligible energy difference between the two phases and the pronounced band sensitivity to exchange splitting support that $Mo_4PtGa_{17}$ is near a magnetic instability, consistent with experimental observations. Moreover, the nonmagnetic phase of $Mo_4PtGa_{17}$ breaks inversion symmetry but exhibits several mirror planes. The Kramers degeneracy splits away from time-reversal-invariant k-points (L, Γ and F) but cannot form Weyl points, because L, Γ and F lie inside a mirror plane that forbids monopoles. Instead, the Kramers degeneracy extends along some path that connects two neighboring time-reversal-invariant *k*-points, presenting a Kramers nodal line[44], making $Mo_4PtGa_{17}$ a topological nodal line metal candidate.

To investigate how local correlation effects influence the low-energy electronic structure, we performed DFT+U calculations with various U values. As shown in Fig. 4c, the low-energy DOS enhancement persists over a moderate on-site correlation from U = 0 - 4 eV, although its peak positions and spectral weights evolve quantitatively. It suggests the absence of localized bands as one would expect for conventional *f*-electron heavy-fermion compounds. Extended Data Fig. S9 illustrates the orbital-projected band structure and DOS of Mo1/Mo2-*d* and Ga-*p* orbitals when U = 0 and 2 eV. It can be clearly observed that the Mo-4*d*-derived features persist under U, confirming the robustness of the flat-band and vHS features, indicating that these are intrinsic nature originating from the crystal structure, not a U-driven behavior. However, to conduct fully quantitative treatment of the correlated state, dynamical approaches such as DFT+DMFT and/or ARPES/quantum oscillations are required.

To identify the lattice origin of the flat bands associated with the near-$E_F$ vHS feature, we constructed a four-band tight-binding model on the Mo breathing pyrochlore lattice. In the ideal pyrochlore case, equal nearest-neighbor (NN) hoppings produce a twofold flat band (Fig. 4f). This flat band is the three-dimensional analougue of the Kagome flat band and arises from destructive interference of hopping amplitudes on the corner-sharing tetrahedral network.[17] Introducing the breathing anisotropy of $Mo_4PtGa_{17}$ retains the flat band inherited from the parent pyrochlore lattice and separates the dispersive branches (Fig. 4g), consistent with the general flat-band evolution in breathing pyrochlore and kagome lattices.[45] With additional longer-range Mo-Mo hoppings, the flat band acquires a weak dispersion and forms a saddle-shaped band near $E_F$, reproducing the DFT feature responsible for the vHS (Figs. 4e & 4h). The DFT and tight-binding results therefore identify the Mo breathing pyrochlore lattice as the geometric origin of the flat band feature associated with the near-$E_F$ vHS in $Mo_4PtGa_{17}$.

**Discussion:** The heavy-fermion-like normal state of $Mo_4PtGa_{17}$ is rarely observed in *d*-electron materials. Meanwhile, prior *d*-electron heavy-fermion-like systems are generally discussed in different microscopic frameworks. In $LiV_2O_4$, the microscopic origin of the heavy *d*-electron states has been discussed in terms of orbital selectivity and frustrated local-moment physics. In particular, orbital-selective $^{51}V$ NMR identified orbital-dependent local moments, persistent antiferromagnetic correlations and a frustrated spin-liquid component coupled to itinerant electrons.[12] Therefore, $LiV_2O_4$ represents a frustrated *d*-electron heavy-fermion metal in which orbital-selective localized and itinerant degrees of freedom play an essential role. By contrast, in $Mo_4PtGa_{17}$, it does not show clear evidence for localized moments, supported by the small effective moments from high-temperature magnetic susceptibility fitting in Fig. 2a, the lack of magnetic broadening of NMR/NQR lines shown in Extended Data Fig. S6 and the pronounced sensitivity of the near-$E_F$ states to weak spin polarization in Extended Fig. S8b. Instead, DFT+U calculations resolve the near-$E_F$ bands into a strongly hybridized Mo-*d*/Ga-*p* channels whose dispersion and flat-band/vHS features remain nearly unchanged (Fig. 4b and Extended Data Fig. S9), supporting the lack of localized moments and that $Mo_4PtGa_{17}$ is different from $LiV_2O_4$ and canonical *f*-electron heavy-fermion compounds where localized *f* electrons hybridize with conduction electrons.

On the other hand, the combination of heavy-fermion-like correlation and superconductivity in one system is another intriguing but unusual phenomenon in *d*-electron materials. The only close example is Fe-based superconductors, including $AFe_2As_2$ (A = alkali metal)[9] and $YFe_2Ge_2$.[14] These materials are widely discussed as Hund metals, where strong correlations, coherence-incoherence crossover and heavy quasiparticles are associated with sizable Hund's coupling and orbital selectivity of Fe-3*d* states. In this picture, Hund's coupling among Fe-3*d* orbitals produces orbital-dependent mass renormalization, with the Fe $d_{xy}$ orbital generally being more strongly correlated than the others. This framework is different from the scenario proposed for $Mo_4PtGa_{17}$. The present DFT+U results show that the Mo-*d* and Ga-*p* states remain strongly hybridized within a low-energy regime from U = 0 to 2 eV; the low-energy dispersions are nearly unchanged, while vHS-related DOS peaks shift only slightly (Extended Data Fig. S9). The NMR K-$\chi$ relation in Extended Data Fig. S6b does not show a Knight-shift anomaly as reported in the Fe-based systems, which was used to support the incoherent-coherent crossover.[46,47]

We therefore propose that $Mo_4PtGa_{17}$ is better described as a new geometrically frustrated Mo-4*d* flat-band/vHS route to heavy-fermion-like superconductivity, in which the geometrically frustrated Mo lattices enhance the DOS, spin susceptibility and quasiparticle mass, rather than a canonical *f*-electron Kondo-lattice or Fe-Hund/orbital-selective mechanism. Additionally, the fact that *s*-wave superconductivity emerges from a heavy-fermion-like and strongly spin-fluctuated normal state also deserves further studies.

## Methods

**Crystal growth of $Mo_4PtGa_{17}$:** Molybdenum powder (Thermo Scientific, -100 mesh, 99.95%), platinum powder (Thermo Scientific, -200+400 mesh, 99.98%), and gallium ingots (Thermo Scientific, 99.98%) were mixed in zirconia crucibles with the ratio of 4:1:50. The crucibles were placed in evacuated quartz tubes and sealed under vacuum of approximately 0.02 Torr after being purged with argon. The tubes were first heated to 1150 °C at a rate of 180 °C/h, maintained at this temperature for 48 hours and slowly cooled to 400 °C at a rate of 2 °C/h. Subsequently, the tubes were taken out, quickly flipped over and centrifuged. The products were soaked in 1 – 3 M HCl solution to remove excess Ga. Crystals with metallic luster and triangular faces can be obtained with a dimension of ~ 250 × 150 × 150 $\mu m^3$, as shown in the inset of Extended Data Fig. S1. Note that most of the yielded crystals using this synthetic method tend to form large multidomain clusters. The resulting crystals are stable in air for at least a year.

**Single crystal and powder X-ray diffraction (XRD):** Single crystals of $Mo_4PtGa_{17}$ were characterized by single-crystal XRD to determine the chemical formula and structure. The Bruker D8 QUEST ECO diffractometer equipped with APEX5 software and Mo-source radiation ($\lambda_{K\alpha}$= 0.71073 Å) was operated for single crystal measurements at room temperature. The crystals were soaked in glycerol and then mounted on a Kapton loop. The single-crystal data acquisition was determined by the Bruker SMART software, and corrections for Lorentz and polarization effects were also included. The numerical absorption correction based on crystal-face-indexing was performed by XPREP. The direct method and full-matrix least-squares on $F^2$ procedure within the SHELXTL package would be implemented to solve the single-crystal structure and chemical composition.[48,49]

Powder XRD measurements were utilized to determine the phase purity for $Mo_4PtGa_{17}$. A Bruker D2 PHASER with Cu Kα radiation and a LynxEye-XE detector was employed. The resulting powder XRD patterns were fitted by the Rietveld method using Fullprof Suite. Crystal structures determined by single crystal XRD were used to obtain the calculated powder XRD patterns and to fit the observed patterns.[50,51]

**Scanning electron microscopy (SEM):**

A Ziess Sigma 500 VP SEM with Oxford Aztec X-EDS was used with an electron beam energy of 20 kV to obtain pictures of $Mo_4PtGa_{17}$ crystals.

**Physical property measurement:** Above 2 K, the Quantum Design Dynacool Physical Property (PPMS) was used to measure the DC magnetization from 2 K to 300 K under various applied external magnetic fields using the equipped ACMS II option. Field-dependent magnetization data were collected in the range of -9 to 9 T under different temperatures. Temperature-dependent of magnetic susceptibility for $Mo_4PtGa_{17}$ was measured from 2 to 300 K under different external fields. Zero field cool (ZFC) and field cool (FC) were both performed. The resistivity measurements were performed in the same equipment by using the four-probe method between 2 to 300 K. Platinum wires were used to connect with the samples by silver epoxy to ensure the ohmic contact. Heat capacity was measured using a standard relaxation method in the PPMS. Heat capacity, electrical resistivity and AC magnetic measurements below 2 K were conducted on a Quantum Design PPMS equipped with Dilution Refrigerator. All the physical properties measurements were conducted on picked as-synthesized crystals and/or clusters of crystals that are naturally grown together. No additional annealing was performed.

**Neutron powder diffraction:** Neutron powder diffraction was performed using HB-2C wide-angle neutron diffractometer ($WAND^2$)[52] at the High Flux Isotope Reactor (HFIR) at Oak Ridge National Laboratory (ORNL). A Ge (113) monochromator was used to select a constant wavelength of 1.486 Å at a take-off angle of 51.5°.

**$^{69,71}$Ga nuclear magnetic resonance (NMR):** NMR and nuclear quadrupole resonance (NQR) measurements of $^{69}$Ga ($I$ = 3/2 , $\gamma_N/2\pi$ = 10.2192 MHz/T, Q = 0.178 barns) and $^{71}$Ga ($I$ = 3/2 , $\gamma_N/2\pi$ = 12.9847 MHz/T, Q =

0.112 barns) nuclei were conducted on a powder sample of $Mo_4PtGa_{17}$ using a laboratory-built phase-coherent spin-echo pulse spectrometer. The NMR spectra were obtained by sweeping $H$ at a fixed frequency of 65.1 MHz, while the NQR spectrum was measured in steps of frequency $f$ by measuring the intensity of the Hahn spin echo. The $^{69}$Ga -NMR $1/T_1$ at the Ga1 site (with $\nu_Q \sim 0$) was measured with a saturation recovery method using free induction decay (FID) signal. $1/T_1$ at each $T$ was determined by fitting the nuclear magnetization $M$ versus time $t$ using the single exponential function $1 - M(t)/M(\infty) \ = e^{-t/T_1}$, where $M(t)$/and $M(\infty)$ are the nuclear magnetization at time $t$ after the saturation and the equilibrium nuclear magnetization at $t \rightarrow \infty$, respectively.

**Electronic structure calculation:** DFT calculations were performed with Vienna ab initio Simulation Package (VASP) using the projector-augmented wave method.[53] We employed the Perdew-Burke-Ernzerhof (PBE) exchange-correlation energy functional[54] with spin-orbit coupling. A plane wave cutoff energy was set to 500 eV. The Brillouin zone was sampled with Γ-centered 9×9×9 mesh for the rhombohedral cell (a=8.1747 Å, α=60.0094°) experimentally obtained. For DFT+U calculations, an on-site interaction $U_{\mathrm{Mo},4d}$ = 0 - 4 eV was applied to the Mo-4$d$ states using Dudarev's DFT+U scheme.[55]

The tight-binding bands in Fig. 4 were obtained from a four-band model on the Mo breathing pyrochlore sublattice. The Hamiltonian was constructed as

$$H_{ij}(\boldsymbol{k}) = \epsilon_0 \delta_{ij} + \sum_{R} t_{ij}(\boldsymbol{R}) e^{i\boldsymbol{k}\cdot(\boldsymbol{r}_j + \boldsymbol{R} - \boldsymbol{r}_i)}$$

where $i$, $j$ label the four Mo sublattices. The hopping amplitudes were grouped according to the Mo-Mo connectivity shown in Fig. 4d. The parameters used in Fig. 4 are $\epsilon_0 = -0.49$ eV, $t_1 = -0.24$ eV, $t_2 = -0.15$ eV, $t_3 = -0.02$ eV, and $t_4 = +0.02$ eV. The ideal pyrochlore limit was calculated using $t_{\mathrm{avg}} = (t_1 + t_2)/2$, and omitted NNN hoppings.

## Acknowledgements

C.W., J.T., and X.G. thank the startup funding from the University of Pittsburgh. Part of the low-temperature resistivity measurements were performed in the Western Pennsylvania Quantum Information Core (RRID:SCR_027582) and services and instruments used in this project were graciously supported, in part, by the University of Pittsburgh. The work at Washington University was supported by the National Science Foundation (NSF) Division of Materials Research Award DMR-2512965. B.Y. acknowledges the financial support by the Israel Science Foundation (ISF: 2974/23) and the Penn State Materials Research Science and Engineering Center for Nanoscale Science (MRSEC) under National Science Foundation award DMR-2011839. The NMR data collection was supported by the U.S. Department of Energy, Office of Basic Energy Sciences, Division of Materials Sciences and Engineering. Ames National Laboratory is operated for the U.S. Department of Energy by Iowa State University under Contract No. DE-AC02-07CH11358. Heat capacity measurement performed at the University of Arizona was supported by the U.S. Department of Energy (DOE), Office of Science, Basic Energy Sciences (BES) under Award DE-SC0025301. B.R.B acknowledges support from the National Science Foundation under Award DGE-2137419. A portion of this research used resources at the High Flux Isotope Reactor, a DOE Office of Science User Facility operated by the Oak Ridge National Laboratory. beam time was allocated to HB-2C WAND$^2$ on proposal number IPTS-36508.1.

## Author Contributions

X.G. conceived the project. C.W. and J.T. grew the samples and performed magnetic properties measurements. X.G. performed the single crystal X-ray diffraction measurement and solved the structure. X.G. conducted heat capacity and resistivity measurements above 2 K and the resistivity measurement below 2 K. B.R.B. and T.K. performed heat capacity measurements below 2 K. S.G. and S.R. performed the resistivity and magnetic properties measurements below 2 K. Q.-P.D., Y.L. and Y.F. conducted and interpreted nuclear magnetic resonance measurements and prepared the draft of relevant section. S.A.C. and M.F. conducted neutron powder diffraction. X.G. analyzed the magnetic, electrical transport and heat capacity data. S.R. and T.K. contributed to the data interpretation. H.B. and B.Y. performed theoretical calculations, interpretation and prepared the draft of theory-related section. The paper was initially prepared by X.G. and all co-authors made comments on the paper.

**Competing Interests**

The authors declare no competing interests.

**Supplementary Information**

The supplementary information includes Supplementary Table S1, Supplementary Discussion S1-S7 and Supplementary Reference.

**Fig. 1 Crystal structure of $Mo_4PtGa_{17}$. a.** Overall crystal structure with coordination environment shown where red/orange, dark blue and light blue spheres represent Mo, Pt and Ga atoms, respectively. **b.** The Mo-Ga coordination type represented by Mo2@$Ga_9$ triply capped trigonal prism. **c.** Breathing pyrochlore sublattice of Mo with Mo1 and Mo2 sites marked in orange and red. Mo1-Mo2 bond lengths are also noted. **d.** Crystal structure of $Mo_4PtGa_{17}$ highlighting the $Mo_4$ clusters.

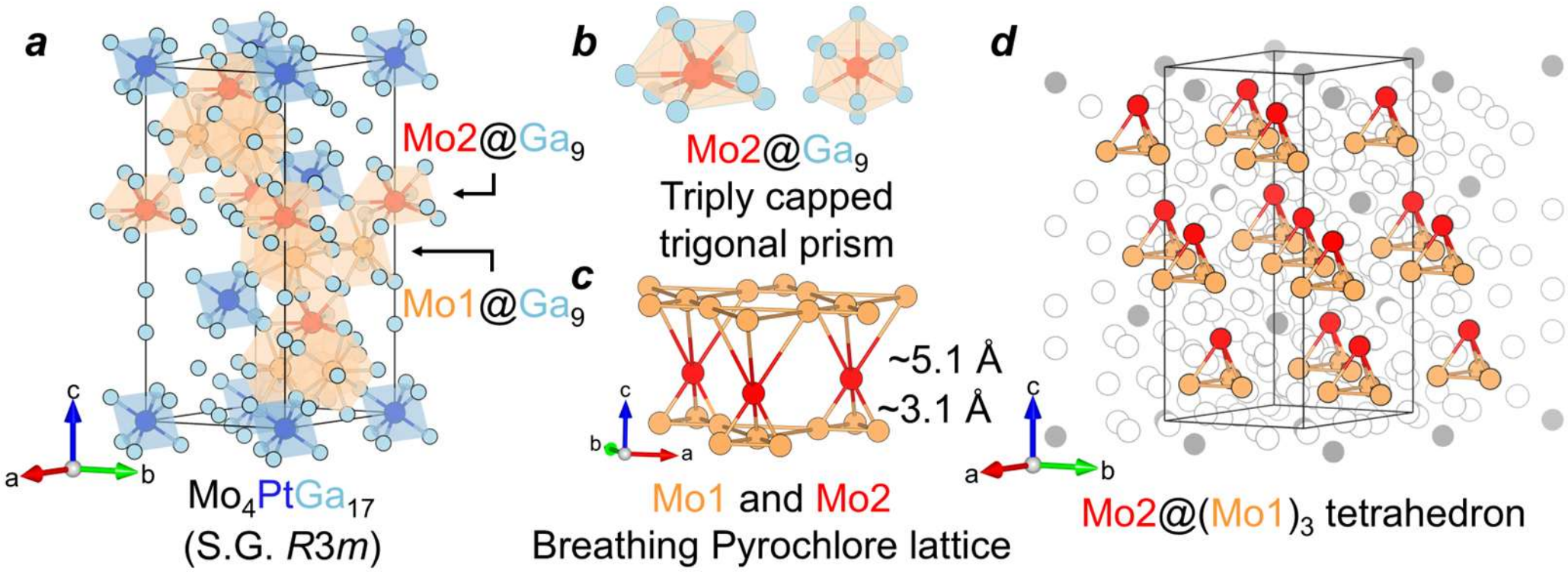

**Fig. 2 Normal-state properties of $Mo_4PtGa_{17}$. a.** Temperature-dependent magnetic susceptibility ($\chi$; solid circle) and its inverse ($\chi^{-1}$; open circle) after subtraction of $\chi_0$ from 2 K to 300 K under magnetic field of 0.5 T. Curie-Weiss fitting is solid line while dashed line is its extrapolation. **b.** Isothermal magnetization under 2 K, 100 K and 300 K. **c. (Main panel)** Temperature-dependent heat capacity ($C_p$) measured under zero magnetic field from 1.9 K to 100 K. **(Left inset)** $C_p/T(T)$ curve. **(Right inset)** $C_p(T)$ with electronic and phononic contributions fitted below 10 K. **d.** Temperature dependence of the spin part of Knight shift ($K_{spin}$). **e. (Main panel)** Temperature dependence of $1/T_1T$. **(Inset)** Temperature dependence of $\alpha$. **f. (Main panel)** Temperature-dependent electrical resistivity ($\rho$) under zero magnetic field from 2 K to 300 K. **(Inset)** The enlarged view of the fitting region.

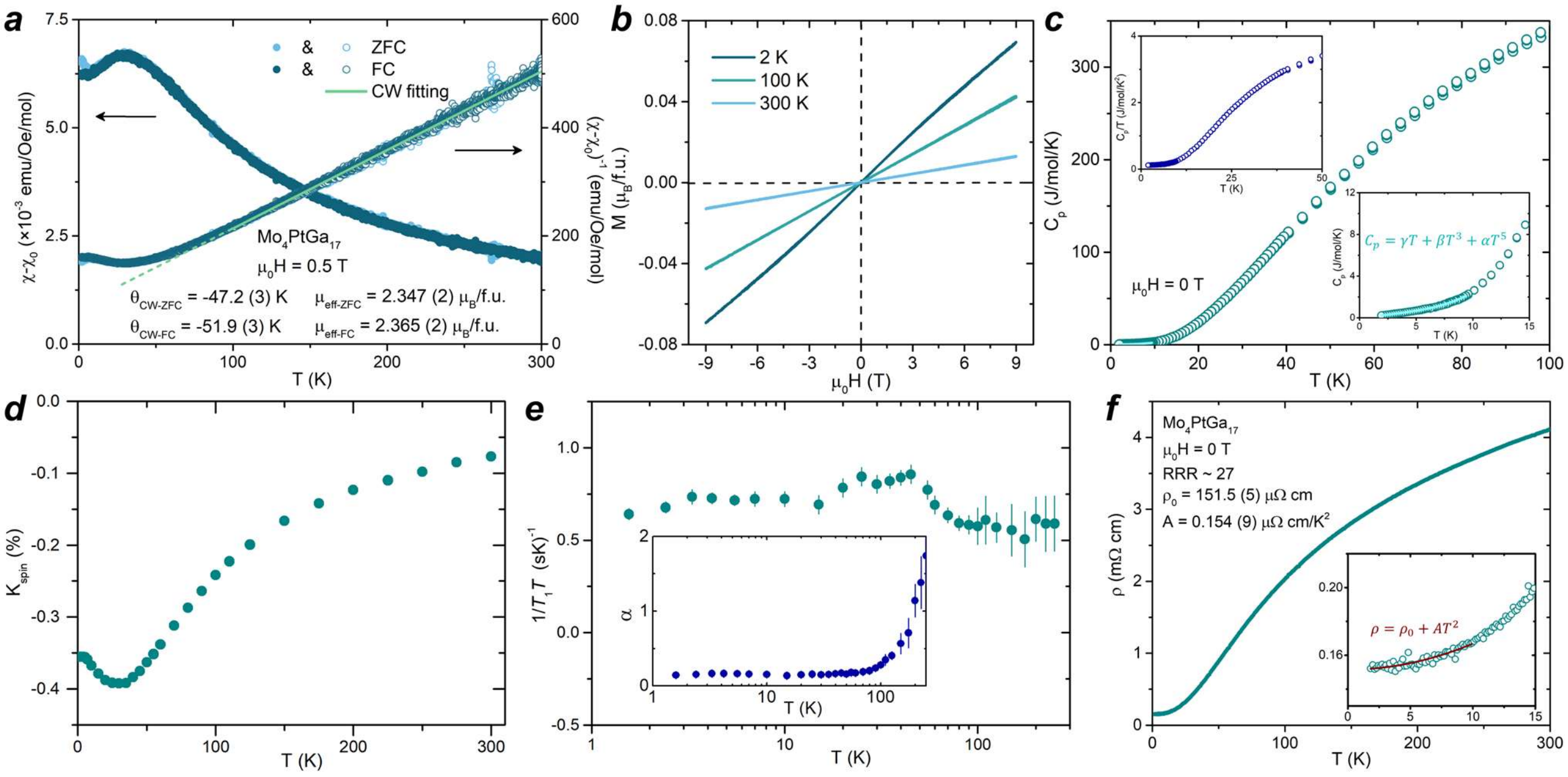

**Fig. 3 Superconducting properties of $Mo_4PtGa_{17}$. a. (Main panel)** Temperature-dependent electrical resistivity (ρ) between 0.1 K and 1.0 K with applied current of 100 μA under various fields. **(Inset)** $\mu_0 H_{c2}$ fitting. **b.** V-I curves between 0.05 K and 0.6 K. The outmost data was measured under 50 mK while the ones next to it were measured under from 300 mK to 600 mK with ΔT = 50 mK. The black arrows indicate the current sweeping direction. **c.** Temperature-dependent AC magnetization data below 1 K showing Meissner effect. A constant background of -71.3 μV is manually subtracted. **d. (Main panel)** Low-temperature heat capacity ($C_p$) between ~0.1 K and 2.0 K under magnetic fields of 0 T and 0.4 T. **(Inset)** $C_p/T(T)$ showing the superconducting heat capacity jump showing the entropy conservation. **e.** $C_p(T)$ fitted using **(Upper panel)** BCS superconducting single-gap function and **(Lower panel)** BCS two-gap function from ~0.1 K to ~0.48 K. The vertical axis is shown on a logarithmic scale. The solid line indicates fitting while dashed line represents extrapolation. **(Inset)** $C_p(T)$ shown in linear scale with two-gap function fitting and extrapolation shown. **f.** $^{69}$Ga NQR results showing **(Main panel)** temperature dependence of $1/T_1$ with the blue solid line showing exponential fitting and **(Inset)** temperature dependence of $1/T_1T$ under multiple magnetic fields.

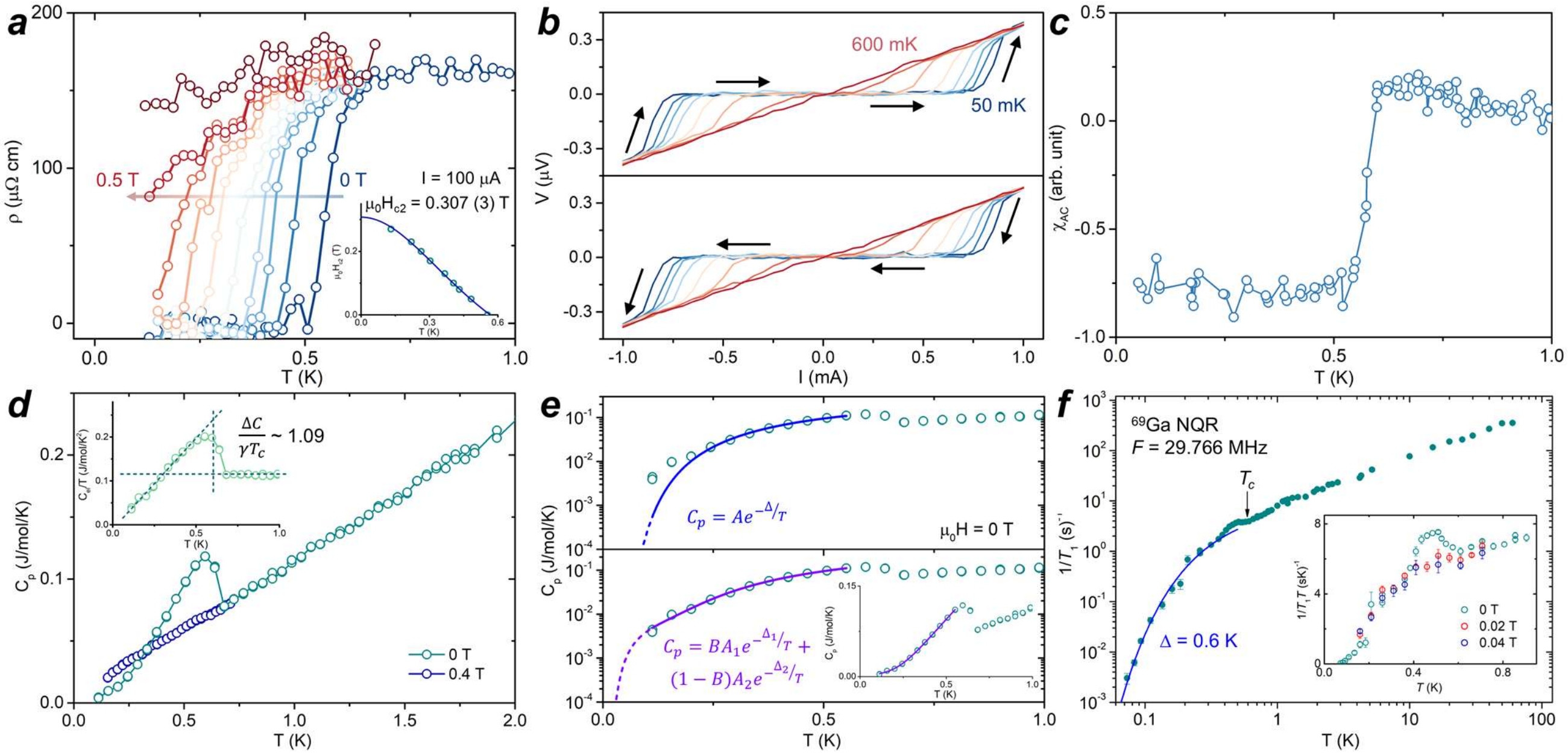

**Fig. 4. Electronic structure of $Mo_4PtGa_{17}$. a.** Nonmagnetic electronic band structure calculated without SOC (gray) and with SOC (red). Arrows highlight saddle-like dispersions expected to generate vHS near $E_F$. The SOC splitting, $\Delta_{SOC}$, is ~116 meV. The right panel shows the corresponding total DOS for the SOC calculation only. **b.** Orbital-projected nonmagnetic band structure and DOS **(Top)** without and **(Bottom)** with SOC. Mo-4*d* and Ga-3*p* are marked in red and cyan, respectively. Pt components are omitted due to the weak contribution near $E_F$. **c.** DOS calculated under various on-site correlation corrections. **d.** $Mo_4$ breathing pyrochlore network and hopping parameters used in the four-band model. The nearest-neighbor (NN) hoppings ($t_1$,$t_2$) distinguish the short and long Mo-Mo bonds, respectively, while the next-nearest-neighbor (NNN) hoppings ($t_3$,$t_4$) connect Mo sites through two-step paths via an intermediate Mo site. **e.** DFT band structure of $Mo_4PtGa_{17}$ without spin-orbit coupling. **f.** Ideal pyrochlore limit with equal NN hoppings ($t_1$=$t_2$), showing the twofold flat band of the parent pyrochlore lattice. **g.** Breathing pyrochlore model with inequivalent NN hoppings ($t_1 \neq t_2$). The breathing anisotropy retains the flat band inherited from the parent pyrochlore lattice and separates the dispersive branches. **h.** Breathing pyrochlore model including NNN hoppings. The longer-range hoppings give the flat band a weak dispersion and bring it into close correspondence with the DFT low-energy band feature associated with the near-$E_F$ DOS enhancement.

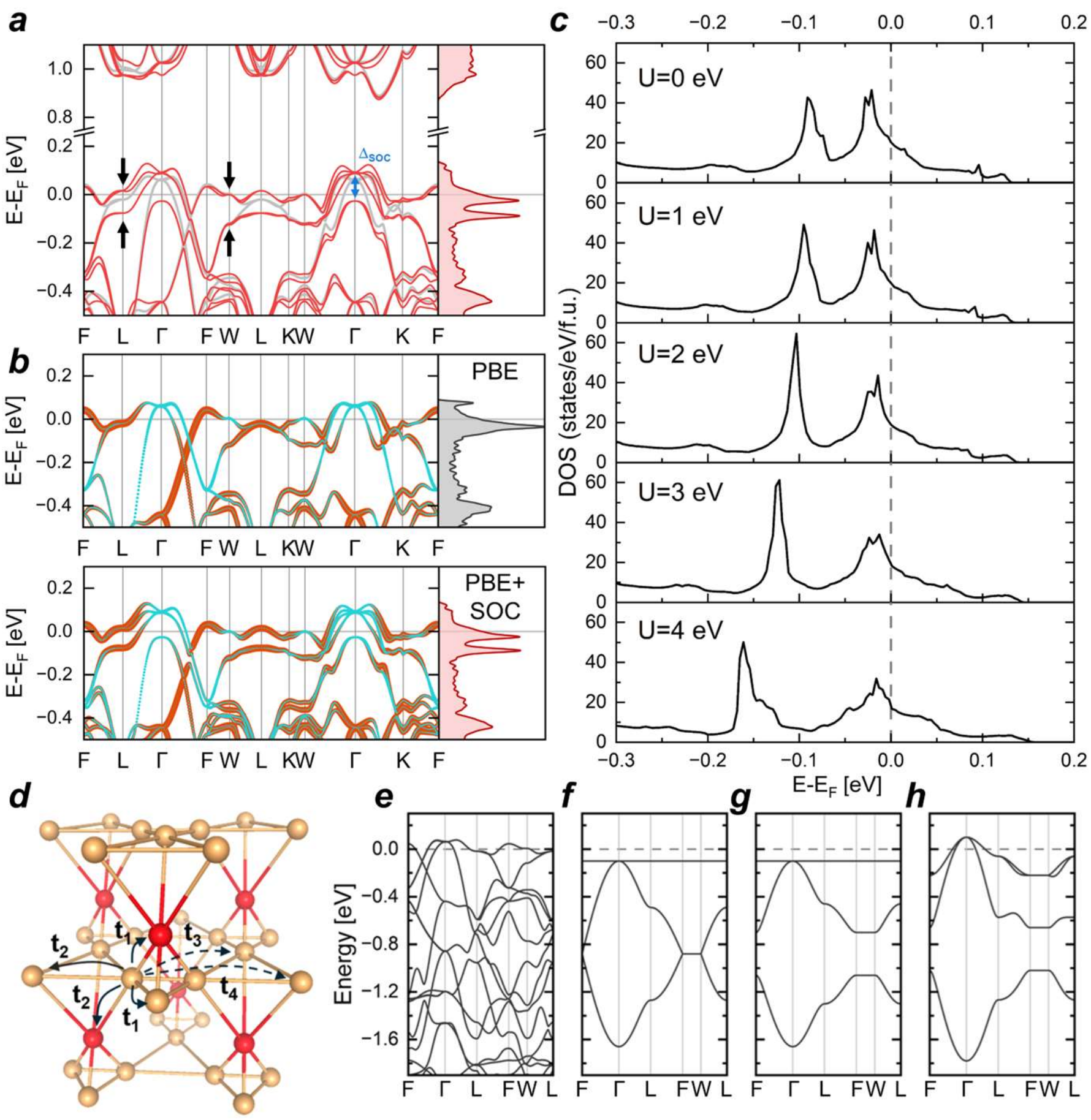

**Extended Data Table S1.** Data collection and refinement statistics for $Mo_4PtGa_{17}$ at 210 (2) K.

| Refined Formula | $Mo_4PtGa_{17}$ |
|---|---|
| Temperature (K) | 210(2) |
| F.W. (g/mol) | 1764.09 |
| Space group; Z | *R3m*; 3 |
| *a*(Å) | 8.189 (1) |
| *c*(Å) | 20.060 (4) |
| V (Å$^3$) | 1164.9 (4) |
| θ range (°) | 3.046 - 33.000 |
| No. reflections; $R_{int}$ | 12007; 0.0992 |
| No. independent reflections | 1171 |
| No. parameters | 51 |
| $R_1$: $\omega R_2$ ($I$>2δ($I$)) | 0.0289; 0.0452 |
| Goodness of fit | 0.978 |
| Diffraction peak and hole (e$^-$/ Å$^3$) | 1.858; -2.212 |
| Absolute structure (Flack) parameter | 0.019(14) |

**Extended Data Table S2.** Atomic coordinates and equivalent isotropic displacement parameters for $Mo_4PtGa_{17}$ at 210 (2) K. ($U_{eq}$ is defined as one-third of the trace of the orthogonalized $U_{ij}$ tensor (Å$^2$))

| **Atom** | **Wyck.** | **Occ.** | ***x*** | ***y*** | ***z*** | $U_{eq}$ |
|---|---|---|---|---|---|---|
| Pt1 | 3*a* | 1 | 0 | 0 | 0.0000(1) | 0.0043(2) |
| Mo1 | 9*b* | 1 | 0.4597(1) | 0.5403(1) | 0.1351(1) | 0.0044(2) |
| Mo2 | 3*a* | 1 | 0 | 0 | 0.5948(1) | 0.0046(4) |
| Ga1 | 3*a* | 1 | 0 | 0 | 0.2499(2) | 0.0064(5) |
| Ga2 | 3*a* | 1 | 0 | 0 | 0.3694(2) | 0.0071(5) |
| Ga3 | 9*b* | 1 | 0.5256(1) | 0.4744(1) | 0.2627(1) | 0.0085(4) |
| Ga4 | 9*b* | 1 | 0.14100(1) | 0.8590(1) | 0.0705(1) | 0.0084(4) |
| Ga5 | 9*b* | 1 | 0.7890(1) | 0.2110(1) | 0.1444(1) | 0.0086(3) |
| Ga6 | 9*b* | 1 | 0.5444(1) | 0.4556(1) | 0.0223(1) | 0.0085(3) |
| Ga7 | 9*b* | 1 | 0.1593(1) | 0.8407(1) | 0.2102(1) | 0.0070(3) |

**Extended Data Table S3.** Anisotropic thermal displacement parameters for $Mo_4PtGa_{17}$.

| Atom | U11 | U22 | U33 | U12 | U13 | U23 |
|---|---|---|---|---|---|---|
| Pt1 | 0.0049(2) | 0.0049(2) | 0.0031(4) | 0.0025(1) | 0 | 0 |
| Mo1 | 0.0047(4) | 0.0047(4) | 0.0039(5) | 0.0023(5) | 0.0001(2) | -0.0001(2) |
| Mo2 | 0.0052(6) | 0.0052(6) | 0.004(1) | 0.0026(3) | 0 | 0 |
| Ga1 | 0.0077(8) | 0.0077(8) | 0.004(1) | 0.0038(4) | 0 | 0 |
| Ga2 | 0.0084(8) | 0.0084(8) | 0.005(1) | 0.0042(4) | 0 | 0 |
| Ga3 | 0.0105(6) | 0.0105(6) | 0.007(1) | 0.0067(7) | -0.0009(3) | 0.0009(3) |
| Ga4 | 0.0092(6) | 0.0092(6) | 0.008(1) | 0.0056(6) | -0.0014(3) | 0.0014(3) |
| Ga5 | 0.0061(5) | 0.0061(5) | 0.010(1) | 0.0006(6) | 0.0000(3) | 0.0000(3) |
| Ga6 | 0.0103(6) | 0.0103(6) | 0.006(1) | 0 | 0 | 0.0035(2) |
| Ga7 | 0.0071(5) | 0.0071(5) | 0.007(1) | 0 | 0 | 0.004(1) |

**Extended Data Fig. S1. a. (Main panel)** Powder XRD pattern of $Mo_4PtGa_{17}$ with Rietveld refinement in red. **(Inset)** SEM picture of a $Mo_4PtGa_{17}$ crystal. **b.** Neutron powder diffraction and calculated patterns under 2.7 K, 10 K and 40 K.

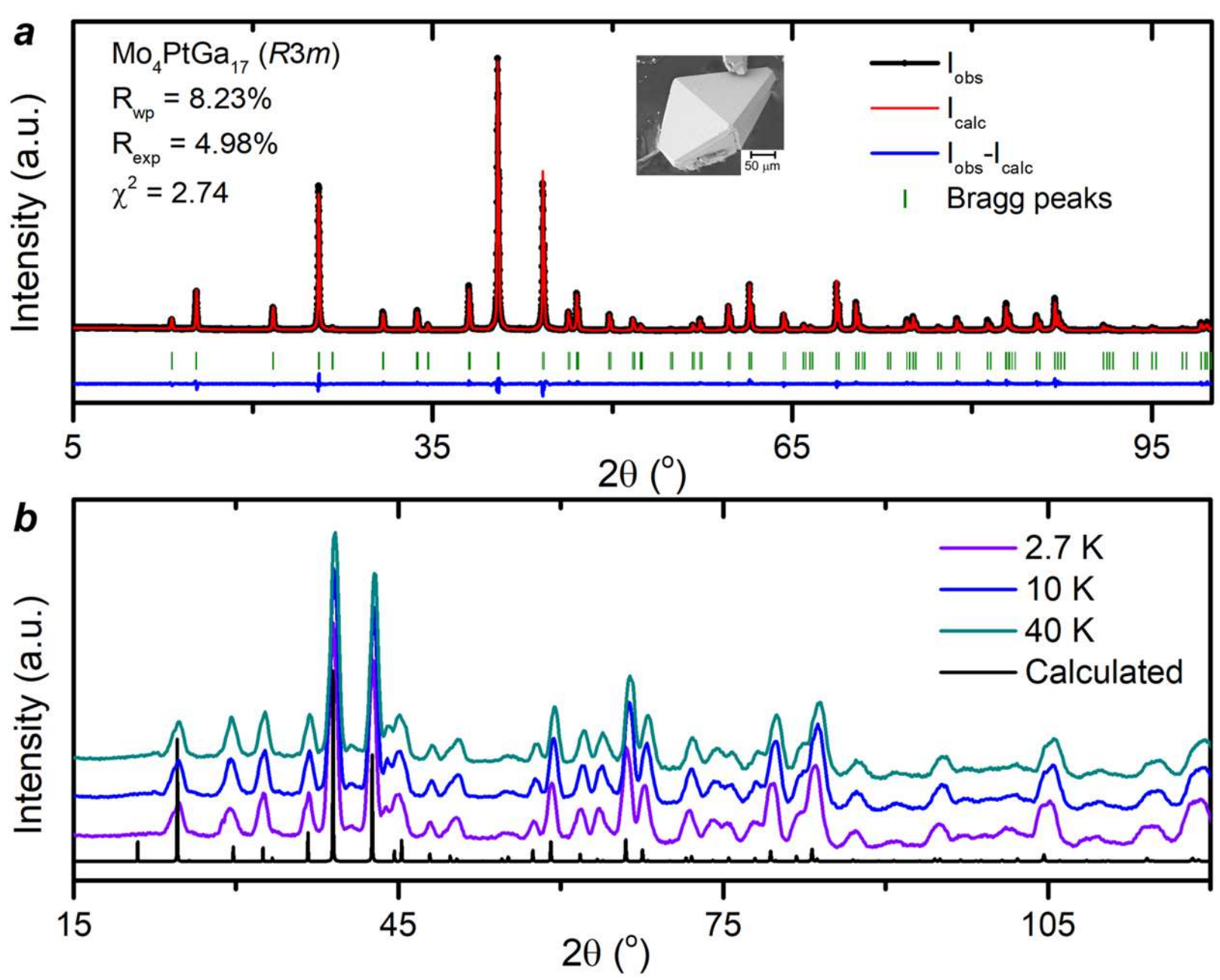

**Extended Data Fig. S2.** Temperature-dependent AC magnetic susceptibility ($\chi_{AC}$') under various frequencies including 500 Hz, 620 Hz, 767 Hz, 950 Hz, 1177 Hz, 1457 Hz, 1804 Hz, 2235 Hz, 2768 Hz, 3427 Hz, 4254 Hz, 5264 Hz, 6511 Hz, 8073 Hz and 9984 Hz. The curves are offset to make better comparison. A DC and AC magnetic field of 10 Oe are applied during the measurement with the AC frequency ranging from 500 Hz to 9984 Hz. No obvious field-dependency can be seen for $\chi_{max}$, suggesting the absence of conventional spin-glass state.

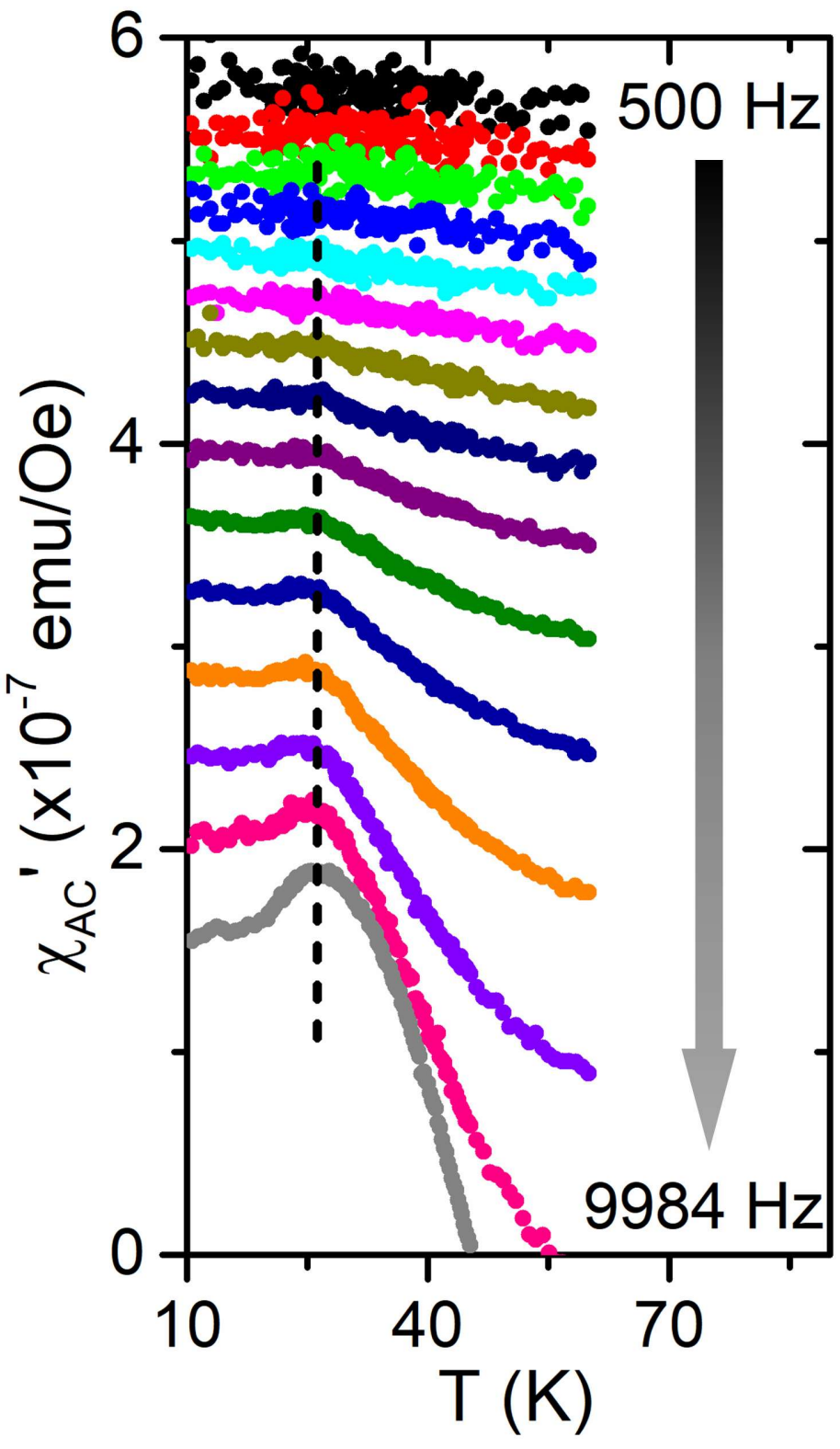

**Extended Data Fig. S3. (Main panel)** Low-temperature moment *vs* temperature curve of the picked crystals of $Mo_4PtGa_{17}$ under an external field of 0.005 T with zero-field cooling protocol. Superconducting transition is seen below ~ 8 K and is attributed to Mo-Pt or Mo-Ga alloy impurities, see Supplementary Information. **(Inset)** Hysteresis loop of the picked crystals under 2 K. Superconducting signal is fully suppressed above ~700 Oe.

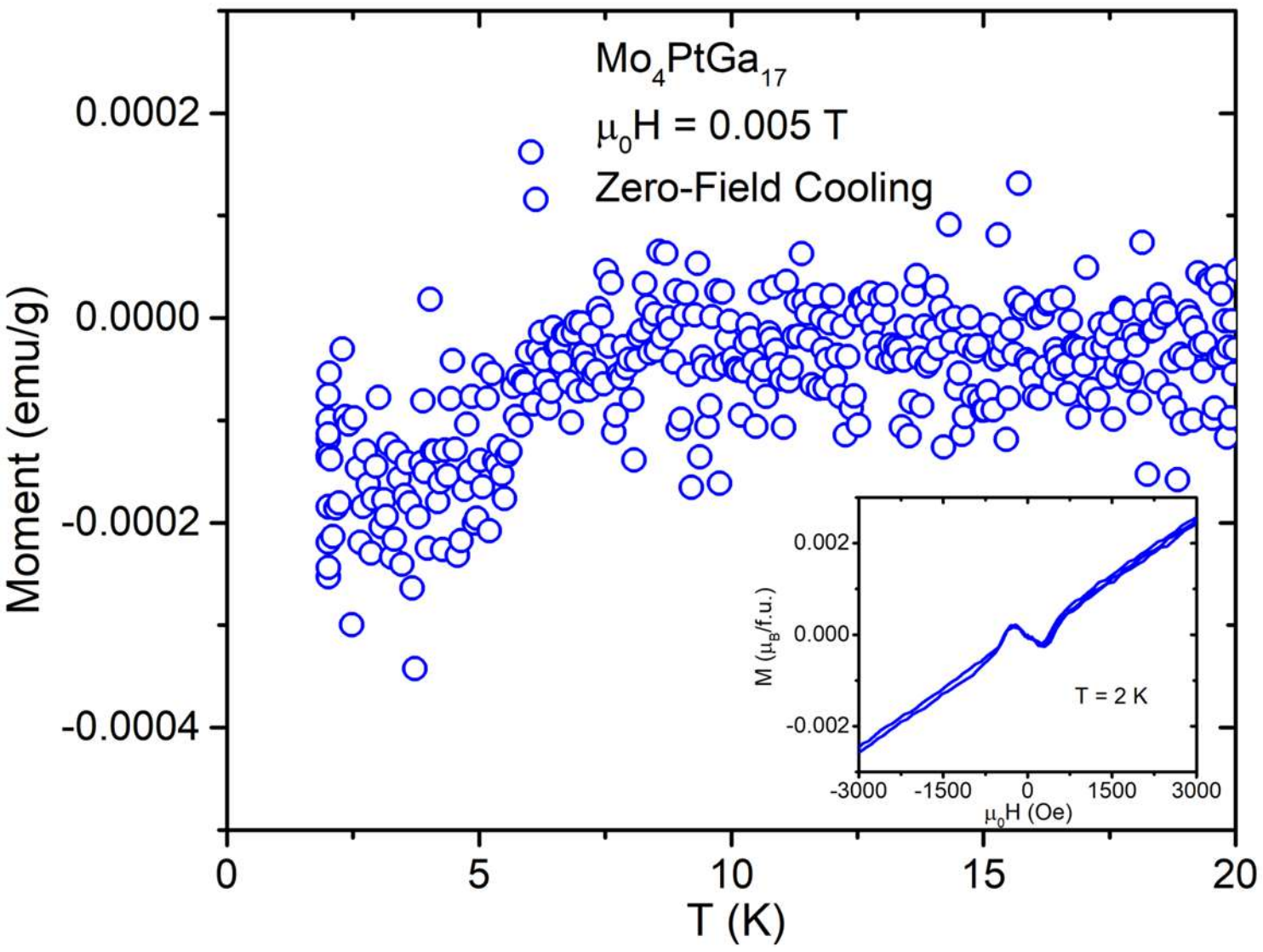

**Extended Data Fig. S4. (Main panel)** Temperature-dependent heat capacity ($C_p$) of $Mo_4PtGa_{17}$ under zero applied magnetic field from 1.9 K to 100 K. The red line represents fitting using electronic term + Debye model from **a.** 1.9 K to 300 K and **b.** 20 K to 300 K. **(Inset)** Enlarged view of low-temperature region where the model fails to describe $C_p$ of $Mo_4PtGa_{17}$. The fitting line and its extrapolation are shown as solid and dashed red lines.

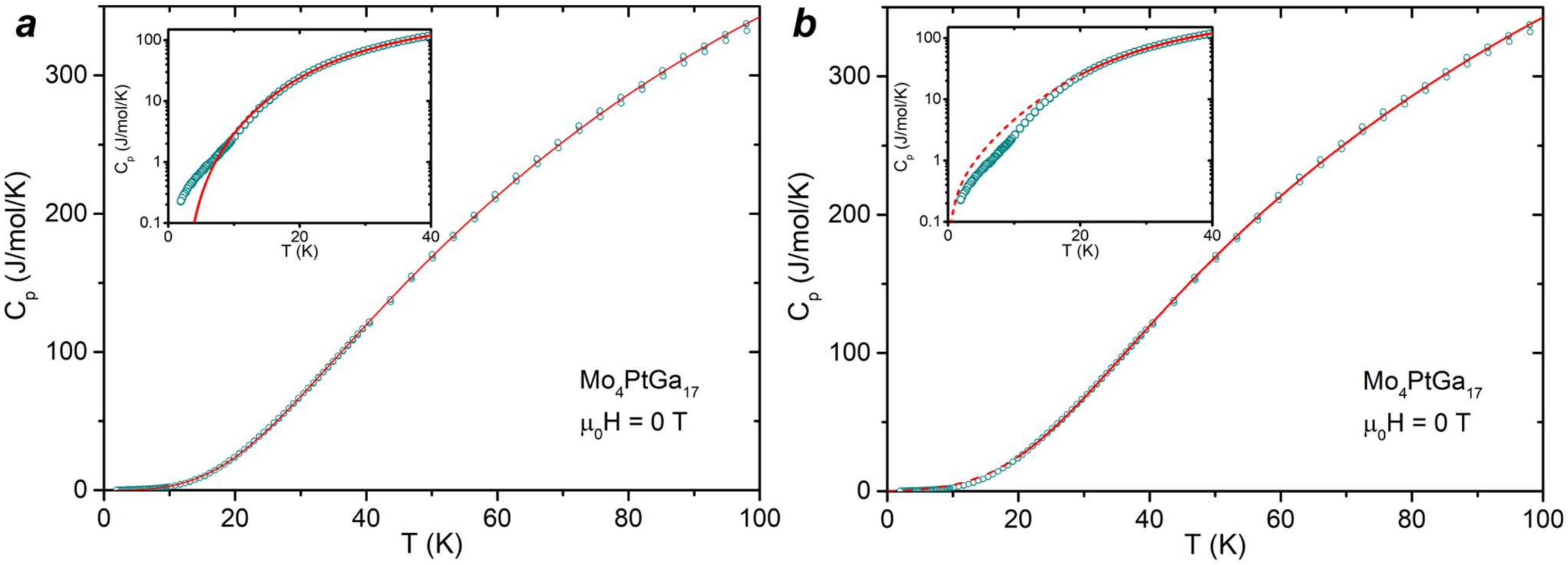

**Extended Data Fig. S5.** $C_p/T(T^2)$ under magnetic fields of 0 T, 5 T, 9 T. The black dashed line is a guide to the eye visualizing the low-temperature linear relation for which the intercept on the y axis does not vary with magnetic fields.

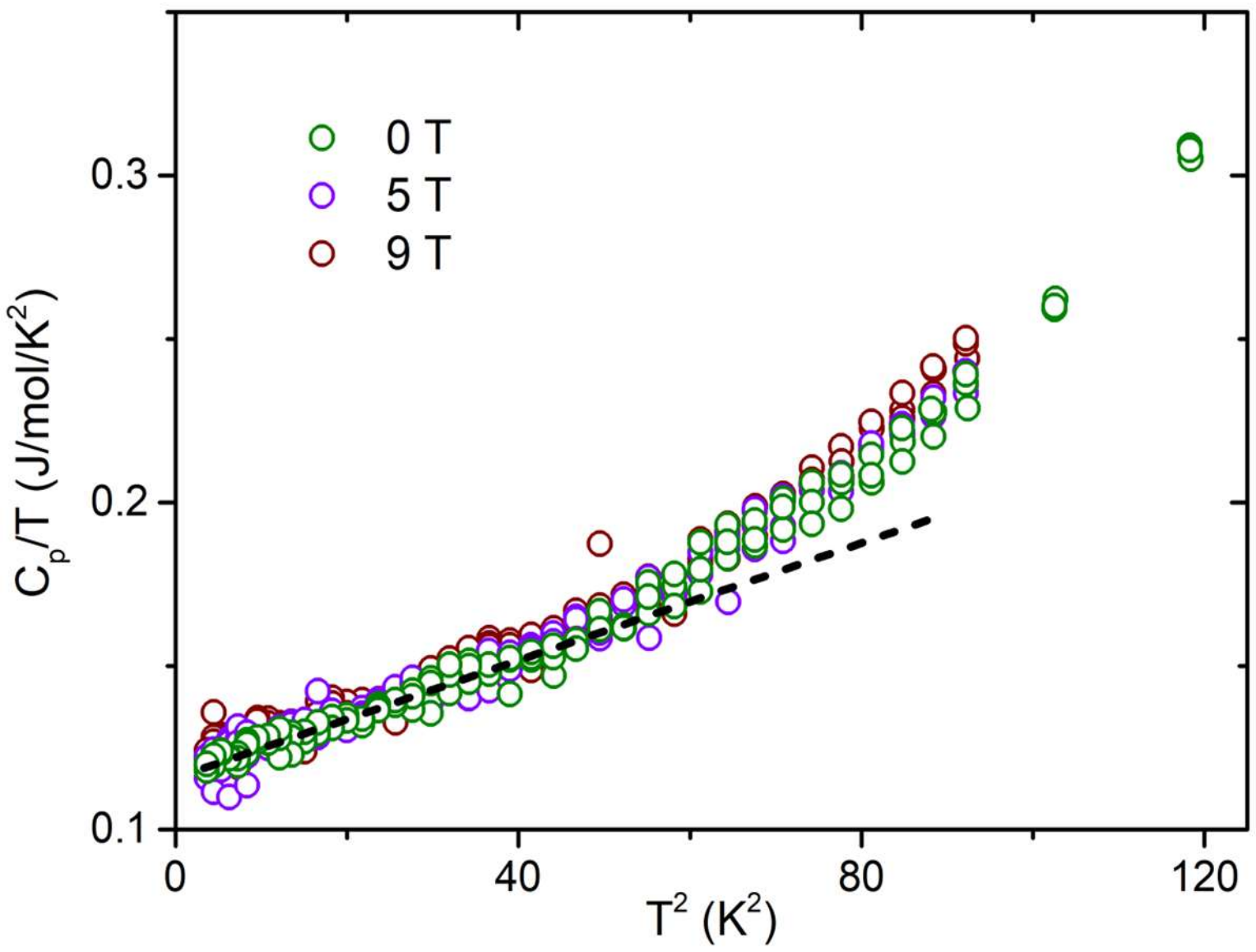

**Extended Data Fig. S6. a.** $^{69,71}$Ga NMR spectrum measured at 4.2 K and *f* = 65.1 MHz (black line). The red curve is the total sum of the simulated spectra shown in (i)-(iv). The inset shows the nuclear quadrupole resonance (NQR) spectrum at 4.2 K. The higher and lower frequency lines for each pair correspond to $^{69}$Ga and $^{71}$Ga lines, respectively. (i) The simulated spectrum for Ga2 and Ga7 with $\nu_Q$ = 29.8(18.7) MHz for $^{69}$Ga ($^{71}$Ga) and $\eta$ = 0. (ii) The simulated spectrum for Ga5 and Ga6 with $\nu_Q$ = 40.2(25.3) MHz for $^{69}$Ga ($^{71}$Ga) and $\eta$ = 0.143. (iii) The simulated spectrum for Ga3and Ga4 with $\nu_Q$ = 35.1(22.1) MHz for $^{69}$Ga ($^{71}$Ga) and $\eta$ = 0.836. (iv) The simulated spectrum for Ga1 with $\nu_Q$ = 0 MHz for $^{69}$Ga($^{71}$Ga) and $\eta$ = 0. The spectra shown in the green and orange areas are from the isotopes $^{69}$Ga and $^{71}$Ga, respectively, and an appropriate broadening was introduced. **b.** *K*-$\chi$ plot. The solid line is the fitting result.

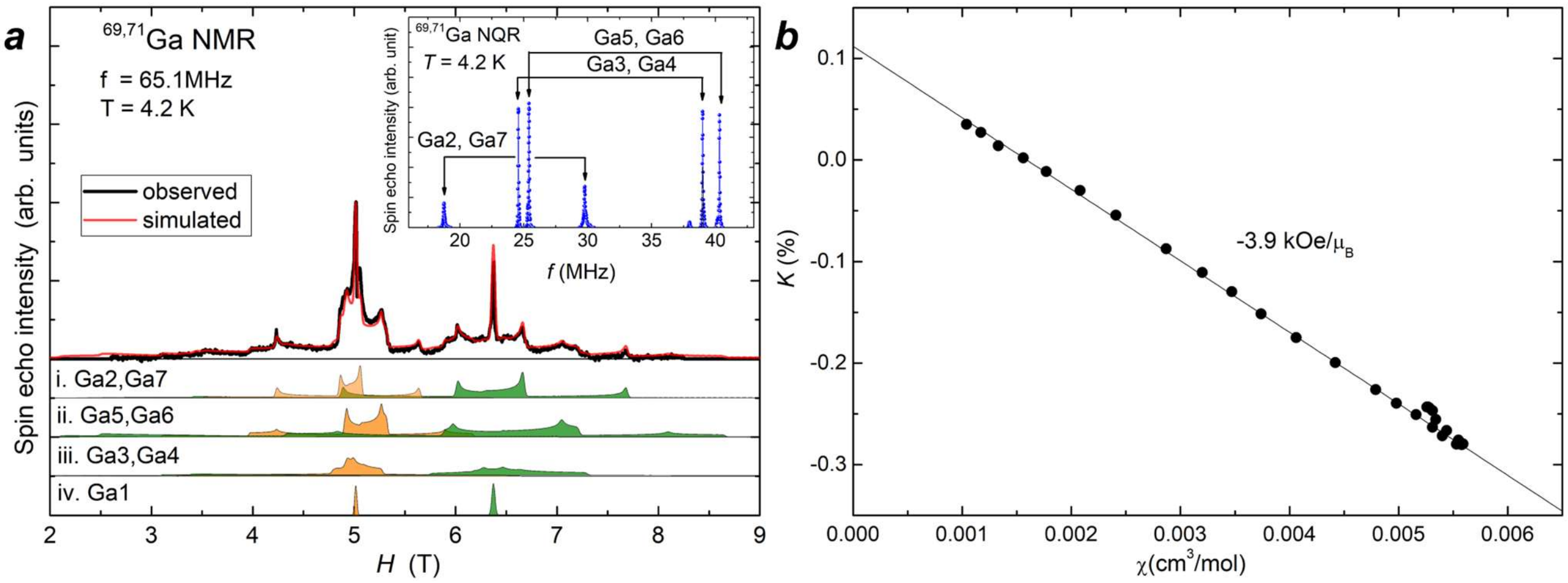

**Extended Data Fig. S7. a.** $\mu_0 H_{c2}$ vs temperarture curve with linear fitting between ~0.37 K and ~0.55 K. **b.** $I_c(T)$ extracted from Fig. 3b.

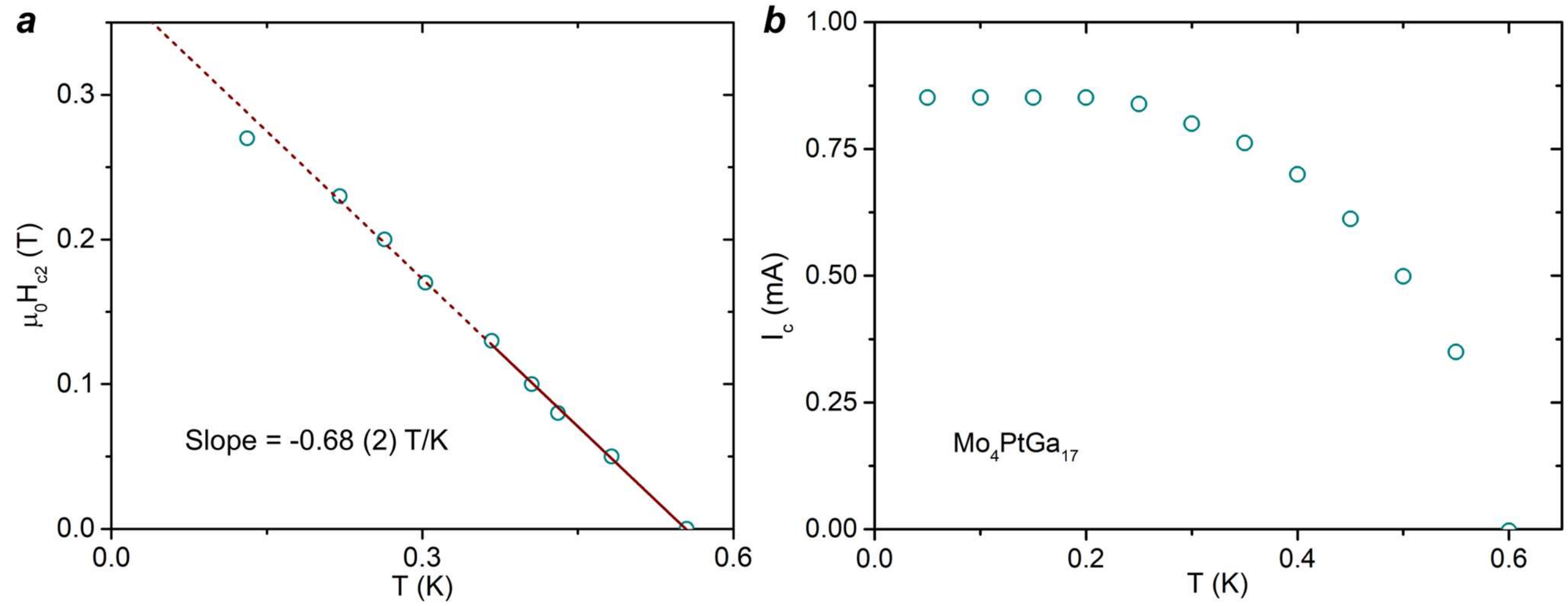

**Extended Data Fig. S8. a.** First Brillouin zone of the primitive rhombohedral cell and high-symmetry *k*-points. Shaded faces are guides to visualize symmetry planes relevant to the chosen momentum cuts. Schematic of the rhombohedral unit cell is included for orientation; Mo1, Mo2, and Pt atoms are shown in orange, red, and blue spheres, respectively. Ga atoms are omitted for clarity. **b.** SOC-included band structures for the nonmagnetic phase (red) and weakly ferromagnetic solutions (blue). The appearance of a small exchange splitting reorganizes the near-EF dispersions and splits nearby spectral features, implying that the low-energy electronic structure is highly sensitive to weak magnetization. The very small total energy difference between two solutions (~10 μeV/u.c.) suggests proximity to a magnetic instability.

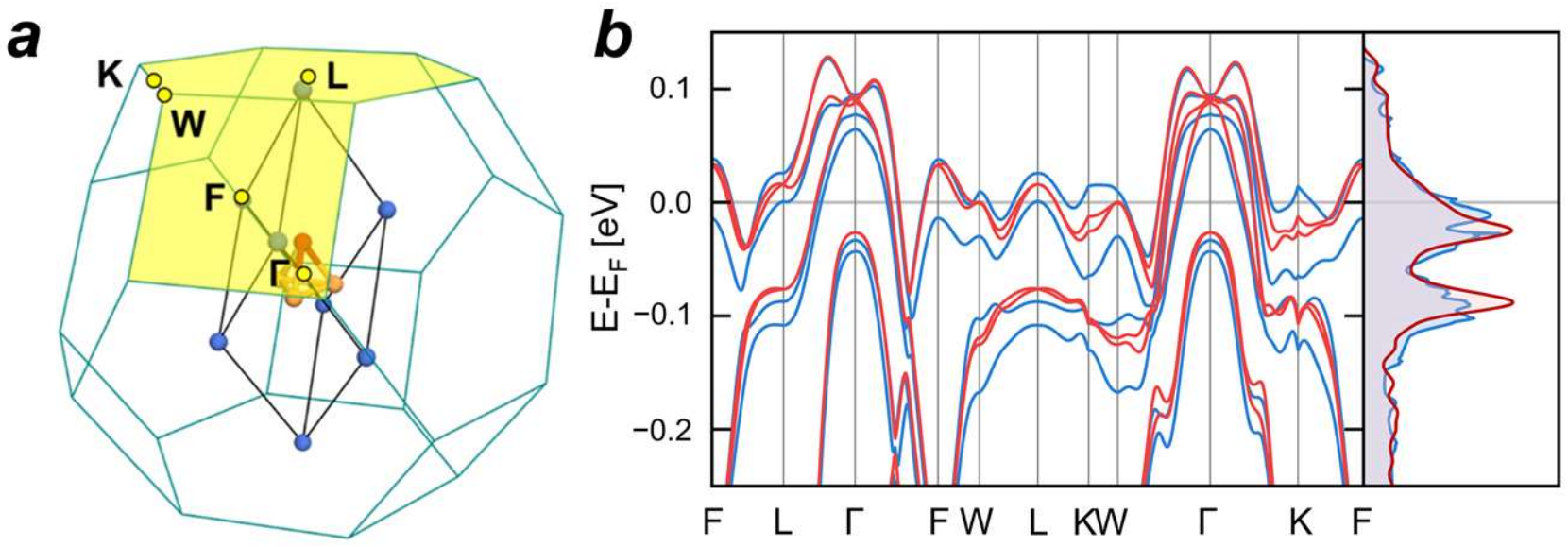

**Extended Data Fig. S9.** On-site interaction $U_{Mo,4d}$ dependence on the species-resolved low-energy electronic structure with SOC. Spectral weight of Mo2-d channel is normalized for visibility and comparison. In DOS, $U_{Mo,4d}$=0 and 2 eV results are shown in black and red lines, respectively.

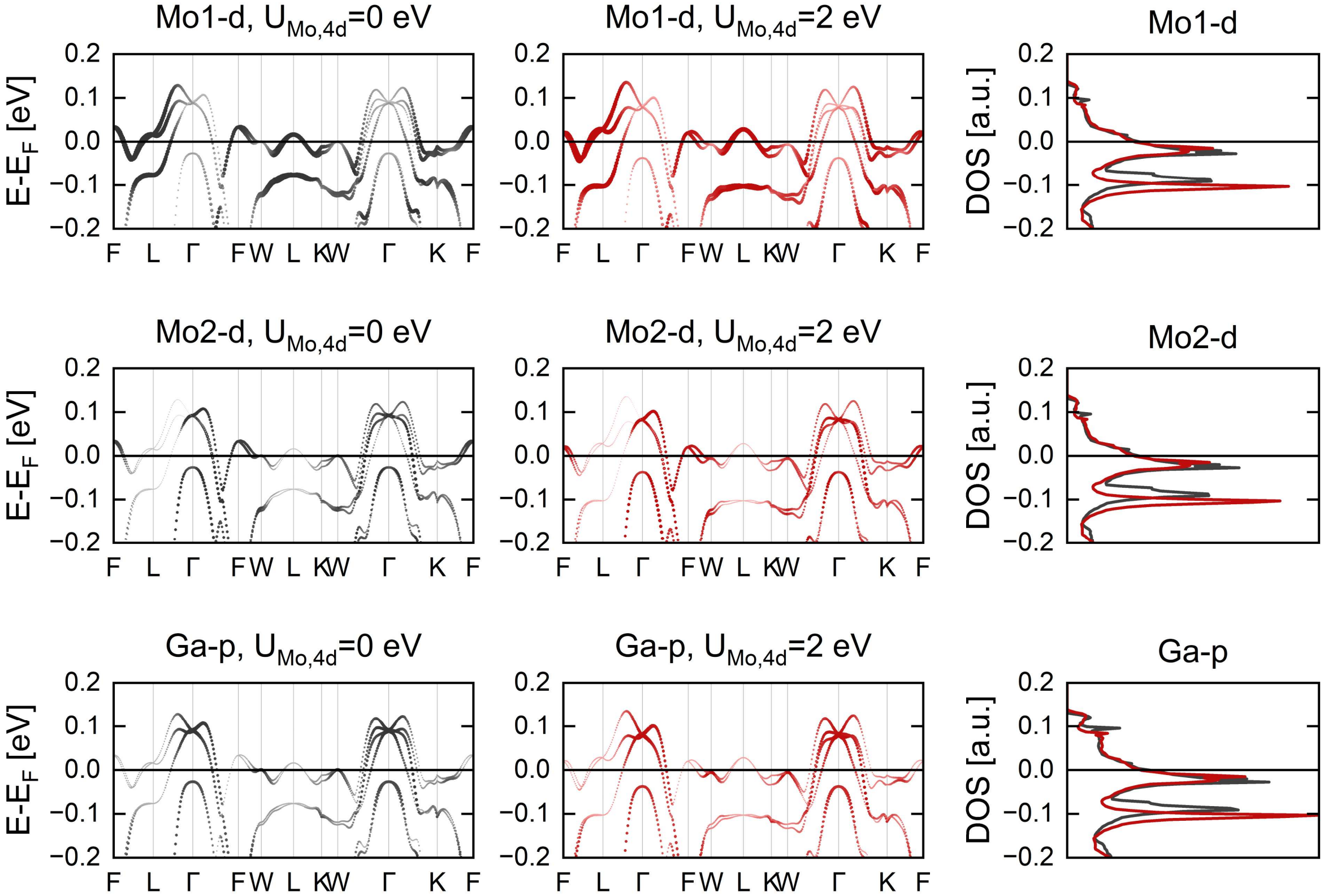

**Supplementary Information: A *d*-Electron Heavy-Fermion-Like Superconductor with Frustration-Induced Flat Bands**

Chaoguo Wang,[1†] Hyeonhu Bae,[2†] Jiaqi Tian,[1†] Qing-Ping Ding,[3] Yongbin Lee,[3] Yuji Furukawa,[3,4] Shannon Gould,[5] Brianna R. Billingsley,[6] Si Athena Chen,[7] Matthias Frontzek,[7] Tai Kong,[6,8] Binghai Yan,[2,9] Sheng Ran[5] and Xin Gui[1]*

[1] Department of Chemistry, University of Pittsburgh, Pittsburgh, PA, 15260, USA

[2] Department of Condensed Matter Physics, Weizmann Institute of Science, Rehovot, 7610001, Israel

[3] Ames National Laboratory, U.S. Department of Energy, Ames, IA, 50011, USA

[4] Department of Physics and Astronomy, Iowa State University, Ames, IA, 50011, USA

[5] Department of Physics, Washington University at St. Louis, St. Louis, MO, 63130, USA

[6] Department of Physics, University of Arizona, Tucson, AZ, 85721, USA

[7] Neutron Scattering Division, Oak Ridge National Laboratory, Oak Ridge, TN, 37831, USA

[8] Department of Chemistry and Biochemistry, University of Arizona, Tucson, AZ, 85721, USA

[9] Department of Physics, The Pennsylvania State University, University Park, PA, 16802, USA

[†]These authors contributed equally to this work.

*Correspondence to: xig75@pitt.edu

**Supplementary Table S1.** Wyckoff position, local symmetry, calculated $\nu_Q$ for $^{69}$Ga($^{71}$Ga) (MHz), observed $\nu_Q$ for $^{69}$Ga($^{71}$Ga) (MHz) at 4.2 K, calculated $\eta$ in $Mo_4PtGa_{17}$. The experimental values of $\nu_Q$ were determined at 4.2K.

| | Wycoff position | local symmetry | cal. $^{69}$Ga($^{71}$Ga) $\nu$Q (MHz) | obs. $^{69}$Ga($^{71}$Ga) $\nu$Q (MHz) | cal. $\eta$ |
|---|---|---|---|---|---|
| Ga1 | 3a | 3m | 0.088(0.055) | ~0(0) | 0 |
| Ga2 | 3a | 3m | 27.2(17.1) | 29.8(18.7) | 0 |
| Ga3 | 9b | .m | 33.25(20.92) | 35.1(22.1) | 0.836 |
| Ga4 | 9b | .m | 33.29(20.94) | 35.1(22.1) | 0.835 |
| Ga5 | 9b | .m | 38.07(23.95) | 40.2(25.3) | 0.142 |
| Ga6 | 9b | .m | 38.08(23.96) | 40.2(25.3) | 0.143 |
| Ga7 | 9b | .m | 27.2(17.1) | 29.8(18.7) | 0.0005 |

**Supplementary Discussion S1.**

**Powder XRD and neutron diffraction:** As mentioned in above, $Mo_4PtGa_{17}$ was synthesized using Ga self-flux method, which, however, only yielded small multidomain crystals shown in the inset of Extended Data Fig. S1a that do not suffice orientation-dependent physical properties measurements. Therefore, powder X-ray diffraction was performed to ensure the sample's purity and homogeneity. It needs to be noted that the powder XRD's sample chamber has a higher temperature (typically 40-50 °C). As can be seen in the main panel of Extended Data Fig. S1a, the results of Rietveld fitting are $R_{wp}$ = 8.23%, $R_{exp}$ = 4.98% and $c^2$ = 2.74. Given the absence of additional peaks, it suggests high purity and high-quality fitting. The fitted lattice parameters are *a* = 8.19253 (4) Å, *c* = 20.0577 (2) Å.

**Supplementary Discussion S2.**

**Subtraction of $c_0$ and diamagnetic background:** The high-temperature c(T) curves are fitted using modified CW law after subtracting a diamagnetic background from the sample holder (~-0.00019 emu at $m_0H$ = 0.5 T)

$$\chi = \chi_0 + \frac{C}{T - \theta_{CW}}$$

where $c_0$ and C are independent of temperature (the latter is related to the effective moment, $m_{eff}$, and the former to the core diamagnetism and temperature independent paramagnetic contributions), and $q_{CW}$ is the CW temperature. The fitted $c_0$ is $-8.179 \times 10^{-4}$ emu/Oe/mol, which is subsequently subtracted from c and result in $c$-$c_0$ in Fig. 2a.

**Supplementary Discussion S3.**

**Superconducting impurities:** As shown in Extended Data Fig. S3, a weak Meissner-like signal is seen below ~ 8 K under an external magnetic field of 0.005 T, which can be suppressed by an external field of ~700 Oe. The trace superconducting impurity can be A15-type Mo-Pt alloys[1] and/or Mo-Ga binary phases[2]. Given the very small estimated volume fraction of this superconducting impurity (< 0.1%), it is believed that the superconducting impurities do not affect the analysis on bulk magnetic properties of $Mo_4PtGa_{17}$.

**Supplementary Discussion S4.**

**Absence of long-range magnetic ordering:** Heat capacity ($C_p$) measurement was conducted between 1.9 K and 100 K under zero external magnetic field. As shown in the main panel of Fig. 2c, no anomaly is seen in the entire measurement range, indicating the lack of any phase transition. The other possibility for $c_{max}$ can be short-range antiferromagnetic ordering if localized moments exist, which typically shows a broad hump in $C_p$(T) or $C_p$/T(T), or both at the corresponding temperatures. However, such behavior is absent in the $C_p$/T(T) curve plotted in the left inset of Fig. 2c, indicating zero or undetectable magnetic entropy change.

**Supplementary Discussion S5.**

**Phonon subtraction from $C_p$:** The total heat capacity of a magnetic material in the paramagnetic region can be interpreted as the sum of electronic plus phononic plus magnonic contributions, as for $C_p = C_{el} + C_{ph} + C_{mag}$. Due to the lack of non-magnetic model of this system, $C_{el}$ is simplified as gT where g is the Sommerfeld coefficient related to electronic contribution to $C_p$. $C_{ph}$ is estimated by the Debye model and thus $C_p$ is represented by the formula

$$C_p = \gamma T + 9nR[C_1(\frac{T}{\theta_{D1}})^3 \int_0^{\theta_{D1}/T} \frac{x^4 e^x}{(e^x - 1)^2} dx + C_2(\frac{T}{\theta_{D2}})^3 \int_0^{\theta_{D2}/T} \frac{x^4 e^x}{(e^x - 1)^2} dx]$$

where n is the number of atoms per formula unit, R is gas constant, $C_1$ and $C_2$ are constants and $C_1 + C_2 = 1$, $q_{D1}$ and $q_{D2}$ are Debye temperatures for two distinct groups of phonons. As shown in Extended Data Fig. S4a, the model failed to describe the low-temperature behavior of $Mo_4PtGa_{17}$ when the fitting range is 2 K to 300 K. The resulted g, $q_1$, $q_2$, $C_1$ and $C_2$ in Extended Data Fig. S4a are -0.023 (13) J/mol/$K^2$, 195 (1) K, 461 (6) K, 0.51 (1) and 0.49 (1), respectively. The negative Sommerfeld coefficient further doubts the validity of the fitting due to the metallic nature of $Mo_4PtGa_{17}$. The best fitting with positive g appears when fitted from 20 K to 100 K, as shown in Extended Data Fig. S4b, with g, $q_1$, $q_2$, $C_1$ and $C_2$ being 0.19 (5) J/mol/$K^2$, 211 (3) K, 561 (27) K, 0.57 (1) and 0.43 (1), respectively. However, the extrapolated curve deviates from the observed heat capacity below ~20 K, likely due to the overestimation of the low-temperature low-energy phonon modes by the Debye model. Therefore, the fitting using the above model is not used in this paper to explain the heat capacity of $Mo_4PtGa_{17}$.

**Supplementary Discussion S6.**

**Nuclear Magnetic Resonance:** Extended Data Fig. S6a shows the $^{69,71}$Ga NMR spectrum measured at 4.2 K, which exhibits a complex line shape. This complexity is due to the presence of crystallographically distinct Ga sites with different values of nuclear quadrupolar interactions and asymmetric parameter $\eta$ of the electric field gradient (EFG) tensor at each Ga site. The typical spectrum for a nucleus with spin $I$ = 3/2 with Zeeman ($\mathcal{H}_{\mathrm{z}}$ ) and quadrupolar ($\mathcal{H}_{\mathrm{Q}}$) interactions can be described by the following nuclear spin Hamiltonian[3] which produces a spectrum with a central transition line flanked by one satellite line on both sides when $\mathcal{H}_{\mathrm{z}}$ is much greater than $\mathcal{H}_{\mathrm{Q}}$.

$$\mathcal{H} = \mathcal{H}_{\mathrm{z}} + \mathcal{H}_{\mathrm{Q}}$$

(1)

where

$$\mathcal{H}_{\mathrm{z}} = -\gamma_{\mathrm{N}}\hbar(1 + K)\mathbf{H} \cdot \mathrm{I}$$

(2)

and

$$\mathcal{H}_{\mathrm{Q}} = \frac{h\nu_{\mathrm{Q}}}{6}(3I_Z^2 - I^2 + \frac{1}{2}\eta(I_+^2 + I_-^2))$$

(3)

Here $H$ is external field, $\hbar$ is Planck's constant (h) divided by 2π, and $K$ represents the NMR shift. The nuclear quadrupole frequency for $I$ = 3/2 nuclei is given by $\nu_Q = eQV_{ZZ}/2h$, where $Q$ is the nuclear quadrupole moment and $V_{ZZ}$ is the maximum electric field gradient (EFG) at nuclear sites. $\eta$ is the asymmetry parameter of EFG defined by $\frac{V_{xx} - V_{yy}}{V_{zz}}$, with $|V_{ZZ}| \geq |V_{YY}| \geq |V_{XX}|$. Since there are two isotopes ($^{69}$Ga and $^{71}$Ga) and seven different Ga sites in the compound, one may generally expect 42 lines, making the spectrum analysis difficult.

To obtain the information about the values of $\nu_Q$ and $\eta$, we calculated EFG at each Ga site by using DFT calculations. The DFT calculations were performed using a full-potential linear augmented plane wave (FP-LAPW) method, as implemented in wien2k[4]. Experimental lattice parameters described above are adopted and the primitive cell contains one formula unit. The generalized gradient approximation of Perdew, Burke, and Ernzerhof[5] was used for the correlation and exchange potentials. The calculations were performed with 408 $k$-points in the irreducible Brillouin zone (IBZ) and were iterated until charge differences between consecutive iterations are smaller than $1\times10^{-3}$ e and the total energy difference lower than 0.01 mRy. The calculated EFG parameters are summarized in Supplementary Table S1. It turns out that, although there are seven inequivalent Ga sites, one can roughly group them into only four distinct sets of $\nu_Q$ and $\eta$: three pairs of sites and one single site (Ga1). For instance, the $\nu_Q$ and $\eta$ values for Ga3 are nearly identical to those with Ga4. Similarly, Ga5 and Ga6 are found to have similar values, as well as the case of Ga2 and Ga7.

Taking these results into consideration, the observed complex spectrum can be reproduced relatively well, as shown by the red curve, which is the sum of the calculated powder-pattern NMR spectra for all Ga sites. It is noted that the sharp NMR lines with free induction decay (FID) signal are observed around 5 and 6.4 T. These lines are attributed to the NMR lines originating from the Ga1 site with nearly zero quadrupolar interaction. Each calculated powder-pattern NMR spectrum using the experimental $\nu_Q$ values with the calculated $\eta$ values are shown in Extended Data Fig. S6a (i), (ii), (iii) and (iv) for the pairs of Ga2 and Ga7, Ga5 and Ga6, Ga3 and Ga4, and the single Ga1 site, respectively, with two spectra originating from isotopes [$^{69}$Ga in green and $^{71}$Ga in orange]. Here we calculated the powder pattern spectra by fully diagonalizing the nuclear spin Hamiltonian [Eq. 1]. Those parameters also reproduce the NQR spectrum measured under zero magnetic field shown in the inset of Extended Data Fig. S6a. In the NQR spectrum for $I$ =3/2 where $\mathcal{H}$ has only $\mathcal{H}_Q$, one can observe a single transition line at a frequency of $\nu_{NQR} = \nu_Q\sqrt{1+\eta^2/3}$. Utilizing the $\nu_Q$ and $\eta$ values obtained from the analysis of the NMR spectrum, the observed NQR spectrum was well reproduced as shown in the inset of Extended Data Fig. S6a.

Fig. 2d shows the temperature dependence of Knight shift $K$ at the Ga1 site measured by $^{69}$Ga NMR. As described above, we see the FID signal from the Ga1 site. This allows us to measure $K$ with relatively high precision. The temperature dependence of $K$ is consistent with that of $\chi$ [Fig. 2d] where the maximum around 30 K is observed as the minimum in the NMR shift data due to the negative hyperfine coupling constant. The NMR shift has contributions from the temperature-dependent spin part $K_{spin}$ and a temperature-independent orbital part $K_0$: $K = K_{spin} + K_0$. $K_{spin}$ is proportional to the spin susceptibility $\chi_{spin}$ through the hyperfine coupling constant $A_{hf}$ giving $K(T) = K_0 + \frac{A_{hf}}{N_A}\chi_{spin}(T)$, where $N_A$ is Avogadro's number. Extended Data Fig. S6b plots $K$ against $\chi$, respectively, with temperature as an implicit parameter. $K$ varies with the $\chi$ as expected. We estimated the hyperfine coupling constant $A_{hf}$ = (−3.9 ± 0.1) kOe/$\mu_B$ and $K_0$ = 0.011 % by fitting the data (shown by the thick line in the inset). Using this value of $K_0$, the temperature dependence of $K_{spin}$ was extracted and is shown in the main text.

**Supplementary Discussion S7.**

**Superconducting coherence length:** The obtained upper critical field, m$_0$H$_{c2}$ (0), is much smaller than the Pauli limit (~1.04 T), thus indicating that m$_0$H$_{c2}$ (0) is dominated by orbital pair breaking. Therefore, the superconducting coherence length ($\xi$) can be fitted using Werthamer–Helfand–Hohenberg (WHH) formula, $\xi = \sqrt{\phi_0/2\pi H_{c2}}$, where $\phi_0$ is the magnetic flux quantum, and $H_{c2}(0) = -0.693 T_c \frac{dH_{c2}}{dT}|_{T_c}$, where $\frac{dH_{c2}}{dT}|_{T_c}$ is the slope of Extended Data Fig. S7 near $T_c$. Here, $\xi$ is estimated to be ~35 nm.

## Supplementary Reference